\documentclass[showpacs,aps,prd,reprint,superscriptaddress,nofootinbib,longbibliography]{revtex4-2}
\usepackage[colorlinks=true, pdfstartview=FitV, bookmarks=true, bookmarksnumbered=true, breaklinks]{hyperref}
\usepackage[dvipdfmx]{graphicx}
\usepackage{amsmath,amssymb,bm,color,longtable,mathrsfs,amsfonts,slashed,comment,tikz,threeparttable}

\newcommand{\lag}{\mathcal{L}}

\definecolor{DeepPink2}{rgb}{0.932,0.07,0.536}
\definecolor{RoyalBlue1}{rgb}{0.284,0.464,1}
\definecolor{SpringGreen3}{rgb}{0,0.804,0.4}
\definecolor{DarkPastelGreen}{rgb}{0.01,0.69,0.28}
\hypersetup{linkcolor=RoyalBlue1, citecolor=DeepPink2, urlcolor=DarkPastelGreen}

\definecolor{lime}{HTML}{A6CE39}
\DeclareRobustCommand{\orcidicon}{
	\hspace{-3mm}
	\begin{tikzpicture}
	\draw[lime, fill=lime] (0,0) 
	circle [radius=0.16] 
	node[white] {{\fontfamily{qag}\selectfont \tiny ID}};
	\draw[white, fill=white] (-0.0625,0.095) 
	circle [radius=0.007];
	\end{tikzpicture}
	\hspace{-3mm}
}

\foreach \x in {A, ..., Z}{\expandafter\xdef\csname orcid\x\endcsname{\noexpand\href{https://orcid.org/\csname orcidauthor\x\endcsname}
			{\noexpand\orcidicon}}
}

\begin{document}

\title{
Chiral effective theory of scalar and vector diquarks revisited
}

\author{Yonghee~Kim}
\email[]{kimu.ryonhi.phys.kyushu.u.ac.jp@gmail.com}
\affiliation{Department of Physics, Kyushu University, Fukuoka 819-0395, Japan}

\author{Makoto~Oka\orcidB{}}
\email[]{makoto.oka@riken.jp}
\affiliation{Nishina Center for Accelerator-Based Science, RIKEN, Wako 351-0198, Japan}
\affiliation{Advanced Science Research Center, Japan Atomic Energy Agency (JAEA), Tokai 319-1195, Japan}

\author{Kei~Suzuki\orcidC{}}
\email[]{{k.suzuki.2010@th.phys.titech.ac.jp}}
\affiliation{Advanced Science Research Center, Japan Atomic Energy Agency (JAEA), Tokai 319-1195, Japan}

\date{\today}

\begin{abstract}
Chiral effective theory of light diquarks is revisited. 
We construct an effective Lagrangian based on the linear representation of three-flavor chiral symmetry.
Here, we focus on the effect of a chiral and $U(1)_A$ symmetric term originated from an eight-point quark interaction.
From this model, we obtain the mass formulas of scalar, pseudoscalar, vector, and axial-vector diquarks, which also describe the dependence of diquark masses on the spontaneous chiral symmetry breaking and the $U(1)_A$ anomaly.
We regard singly heavy baryons as two-body systems composed of one heavy quark and one diquark and then predict the fate of the mass spectrum and the strong decay widths under chiral symmetry restoration.
\end{abstract}
\maketitle

\section{Introduction}\label{Sec_1}

Hadrons made from quarks and gluons show rich diversity in their energy spectrum
and structures. A major part of the spectrum is occupied by the standard mesons and baryons,
while some exotic hadrons, such as tetraquarks and hadronic molecules, have been
observed and attracted a lot of attention recently.
In exploring the rich structural variety, a few key words are useful. 
From the symmetry viewpoint, chiral symmetry and heavy-quark symmetry happen
to be important for light-quark hadrons and heavy-quark hadrons, respectively.
The light-flavor quarks, $u$, $d$, and $s$, are regarded to obey the properties
predicted by chiral symmetry that is exact when the quark masses are zero.
Chiral symmetry is broken in two ways, by the explicit quark masses and by 
the quark condensate in the vacuum of 
quantum chromodynamics (QCD). 
The latter, spontaneous chiral symmetry breaking (SCSB)~\cite{Nambu:1961tp,Nambu:1961fr}, results in the appearance of
light Nambu-Goldstone bosons, $\pi$, $K$, and so on, and also in the generation
of large (constituent) masses of quarks.
SCSB plays key roles in the spectroscopy of both the light mesons and the light baryons.
A related symmetry is the axial U(1) symmetry, which is anomalously broken~\cite{Bell:1969ts,Adler:1969gk,tHooft:1976rip,tHooft:1976snw,tHooft:1986ooh} by the 
coupling of the two-gluon operator, $G\tilde G$, with the same quantum number, $J^{P}=0^-$.
The axial anomaly mixes different light flavors and is crucial for the pseudoscalar meson spectrum.

On the other hand, the heavy-quark symmetry~\cite{Isgur:1989vq} is the symmetry among the heavy quarks, $c$ and $b$.
For the hadron spectroscopy, it is important to realize that the heavy-quark spin is decoupled from
the spin (and orbital angular momentum) of light quarks. This is because the spin-dependent 
coupling of a heavy quark and a gluon is suppressed by $1/m_Q$ factor.

Another key word is ``diquark''~\cite{GellMann:1964nj, Ida:1966ev, Lichtenberg:1967zz, Lichtenberg:1967, Souza:1967rms, Lichtenberg:1968zz, Carroll:1969ty, Lichtenberg:1981pp} (see Refs.~\cite{RevModPhys.65.1199, JAFFE20051} for reviews).
The diquark is a product of two quarks either in color $\bar{\bm 3}$ or $\bm 6$ 
produced by an interquark attraction at short distances. 
Diquarks can be a component of hadrons that are in total color singlet.
The light diquark with color $\bar{\bm 3}$, flavor $\bar{\bm 3}$ ($ud$, $ds$, or $sd$), and spin 0 is called scalar (S) diquark and is 
supposed to have the lightest mass, which is supported by lattice QCD simulations~\cite{Hess:1998sd,Orginos:2005vr,Alexandrou:2006cq,Babich:2007ah,DeGrand:2007vu,Green:2010vc,ChinQCD,Watanabe:2021nwe,Francis:2021vrr}.
The simplest color singlet system containing the S diquark is the singly heavy baryons $Q qq$, such as $\Lambda_Q$ and $\Xi_Q$, where the light quarks, $qq$, form the S diquark, and the diquark correlation can characterize the baryon mass spectrum and can be reflected in the decay properties of baryons (e.g., see Refs.~\cite{Lichtenberg:1975ap, Lichtenberg:1982jp, Lichtenberg:1982nm, Fleck:1988vm, Ebert:1995fp, Ebert:2007nw, Kim:2011ut, Ebert:2011kk, Chen:2014nyo, Jido:2016yuv, Lu:2016ctt, Kumakawa:2017ffl, CETdiquark, scalar, Dmitrasinovic:2020wye, Kawakami:2020sxd, Suenaga:2021qri, Kim:2021ywp, Suenaga:2022ajn,Kumakawa:2021ujz,Garcia-Tecocoatzi:2022zrf,Kim:2022pyq,Suenaga:2023tcy,Jakhad:2023mni,Garcia-Tecocoatzi:2023btk,takada,Kinutani:2023rcx,Pan:2023hwt,Jakhad:2024fin,Gutierrez-Guerrero:2024him,Suenaga:2024vwr}).\footnote{There are many theoretical works~\cite{Isgur:1991wq, Yan:1992gz, Cho:1992gg, Cho:1994vg, Rosner:1995yu, Pirjol:1997nh,Albertus:2005zy, Cheng:2006dk, Zhong:2007gp, Yasui:2014cwa, Cheng:2015naa, Nagahiro, Chen:2016iyi, Can:2016ksz, Chen:2017sci, Wang:2017hej, Cheng:2017ove, Arifi:2017sac, Wang:2017kfr, Yao:2018jmc, Kawakami:2018olq, Lu:2018utx, Arifi:2018yhr, Wang:2018fjm, Kawakami:2019hpp, Cui:2019dzj, Nieves:2019nol, Wang:2019uaj, Yang:2019cvw, Lu:2019rtg, Chen:2020mpy, Azizi:2020tgh, Yang:2020zrh, Arifi:2020ezz, Yang:2020zjl, Azizi:2020azq, Arifi:2021orx, Arifi:2021wdf, Yang:2021lce, Gong:2021jkb, Kakadiya:2021jtv, Kakadiya:2022zvy,Suh:2022ean, Yanga:2022oog, Yu:2022ymb,Wang:2022dmw,Zhou:2023wrf,Ponkhuha:2024gms,Jakhad:2024fgt,Ortiz-Pacheco:2024qcf} for strong decays of singly heavy baryons.}
In particular, $P$-wave excitations of singly heavy baryons may be interepreted by using the picture of the so-called $\rho$-mode (the excitation between $q$ and $q$) and the $\lambda$-mode (the excitation between $Q$ and $qq$) \cite{Copley:1979wj,Yoshida}, and then the diquark excited states play an essential role.

In this paper, we construct a chiral effective theory of diquarks and study the spectrum and decays of singly heavy baryons 
within the diquark--heavy-quark model.
Diquarks in normal vacuum of QCD acquire masses from various origins, chiral invariant mass, mass induced by the SCSB, and mass induced by axial anomaly.
These contributions are of the same order that the mass spectrum is sensitive to the flavor SU(3) breaking, which appears as the differences among the chiral condensates with different flavor contents.
One of our purposes is to explore possible behaviors of the spectrum and decay patterns under the partial restoration of chiral symmetry. When chiral condensates decrease toward zero at high temperature/density, the spectrum of the diquarks is modified and is reflected to the heavy baryon masses.

This paper is organized as follows.
In Sec.~\ref{Sec:Lagrangian}, we present a chiral effective Lagrangian for the scalar ($0^+$, S)/pseudoscalar ($0^-$, P) diquarks and axial-vector ($1^+$, A)/vector ($1^-$, V) diquarks based on the chiral $SU(3)_R\times SU(3)_L$ symmetry. In this paper, we consider an extra term for the S/P diquark Lagrangian, which was not included in our previous studies~\cite{CETdiquark,scalar}.
In Sec.~\ref{Sec:parameter}, we discuss possible ranges of the parameters in the Lagrangian from the phenomenological viewpoints.
The basic approach is to adjust the diquark masses by fitting masses of the single-heavy baryons, $\Lambda_Q$, $\Sigma_Q$, and $\Xi_Q$. Then the Lagrangian parameters are determined from the diquark masses.
In Sec.~\ref{Sec:restoration}, we consider the behaviors of the diquark masses under the chiral symmetry restoration. 
Chiral restoration is controlled by the condensate parameters by introducing the scaling factor $x$ applied to the chiral condensates.
By changing $x$ from 1 (normal vacuum) to 0 (chiral symmetric vacuum), we study the behaviors of the masses of diquarks as well as those of the heavy baryons.
In Sec.~\ref{Sec:spectra}, we discuss the singly-heavy-baryon spectra within our formulation.
In Sec.~\ref{Sec:decay}, we study the decays, $\Sigma_Q$ to $\Lambda_Q+\pi$, induced by the mixing between S/P diquarks and A/V diquarks.
Due to the changes of the condensates under chiral symmetry restoration, the mass difference of $\Lambda_Q$ and $\Sigma_Q$ changes drastically, and then one sees that the decay widths of $\Sigma_Q$'s are sensitive to the values of the chiral condensates.
The conclusion is given in Sec.~\ref{Sec:summary}.

\section{Chiral Effective Lagrangian for Diquarks} \label{Sec:Lagrangian}

We construct a chiral effective theory of diquarks based on the linear representation of the three-flavor chiral $SU(3)_R\times SU(3)_L$ symmetry.\footnote{The lower-order terms in the chiral effective Lagrangian for the S/P diquarks were first introduced in Refs.~\cite{Hong:2004xn,Giacosa:2006tf} (also see Refs.~\cite{Hong:2004xn,Hong:2004ux} for the nonlinear representation).
The importance of $U(1)_A$ breaking in the $m_{S1}^2$ term was pointed out in Ref.~\cite{CETdiquark} (this term is significant also in the context of the Ginzburg-Landau free energy in color superconductors~\cite{Hatsuda:2006ps}).
The chiral effective Lagrangian for the A/V diquarks was first constructed in Ref.~\cite{Kim:2021ywp}.}
The fundamental degrees of freedom in our Lagrangian are 
the scalar (S), pseudoscalar (P), axial-vector (A), and vector (V) diquarks and the Nambu-Goldstone (NG) boson fields.
In this work, all the diquark fields belong to the color $\bar{\bm 3}$, where the S/P/V diquarks belong to the flavor $\bar{\bm 3}$, and only the A diquark belongs to
 the flavor ${\bm 6}$.

\subsection{Diquark fields and chiral symmetry}
The S and P diquarks are represented by the right-handed diquark $d_{R,i}$ and the left-handed one $d_{L,i}$,
where $i=1,2,3$ is the flavor SU(3) index, and hereafter we will omit the color index.\footnote{We defined $d_{R,i} \equiv \epsilon^{abc} \epsilon_{ijk} (q_{R,j}^{bT} C q_{R,k}^c)$, where $a,b,c$ are the color indices, and $C=i\gamma^0\gamma^2$ is the charge-conjugate matrix.
For the vector diquark field, we defined $d^{\mu}_{ij} \equiv \epsilon^{abc} (q_{L,i}^{bT} C \gamma^\mu q_{R,j}^c)$.}
Under the chiral $SU(3)_R\times SU(3)_L$ transformation, the right-handed or left-handed quark field is transformed as
\begin{eqnarray}
&&q_{R,i} \rightarrow (U_{R})_{ij} q_{R,j}, ~~U_{R} \in SU(3)_{R},\\
&&q_{L,i} \rightarrow (U_{L})_{ij} q_{L,j}, ~~U_{L} \in SU(3)_{L}.
\end{eqnarray}
Then, $d_{R,i}$ and $d_{L,i}$ are transformed as
\begin{eqnarray}
&&d_{R,i} \rightarrow d_{R,j} (U_{R}^\dag)_{ji}, ~~d_{L,i} \rightarrow d_{L,j} (U_{L}^\dag)_{ji}.
\end{eqnarray}
The parity eigenstates are given by 
$d_i^{(+)}=\frac{1}{\sqrt{2}} (d_{R,i} - d_{L,i})$ and
$d_i^{(-)}=\frac{1}{\sqrt{2}}(d_{R,i} + d_{L,i})$ for the S and P diquarks, respectively.
Similarly, the A and V diquarks are represented by $d^{\mu}_{ij}$ which is transformed as
\begin{eqnarray}
&& d^{\mu}_{ij} \rightarrow (U_L)_{im}d^{\mu}_{mn}(U_R^T)_{nj}.
\end{eqnarray}
The parity eigenstates are constructed by
$d^{(+) \mu}_{ij}= \frac{1}{\sqrt{2}}(d^{\mu}_{ij} + d^{\mu}_{ji})$ and
$d^{(-) \mu}_{ij}= \frac{1}{\sqrt{2}}(d^{\mu}_{ij} - d^{\mu}_{ji})$ for the A and V diquarks, respectively.

The NG boson nonet $\Sigma_{ij} = \bar{q}_{R,j} q_{L,i}$ is transformed as 
\begin{eqnarray}
&&\Sigma_{ij} \rightarrow (U_L)_{im} \Sigma_{mn} (U_R^{\dag})_{nj}.
\end{eqnarray}

The chiral symmetry is spontaneously broken by the mean field of $\Sigma$ in the vacuum, which is expressed as
\begin{eqnarray}
\langle\Sigma\rangle={\rm diag}(f_\pi,f_\pi,2f_K-f_\pi)\equiv{\rm diag}(f_\pi,f_\pi,f_s),
\end{eqnarray}
where $f_\pi=92.1$ MeV and $f_K=115.1$ MeV ($f_s= 2f_K-f_{\pi}$) are the decay constants of the pion and the kaon, respectively.

\subsection{Effective Lagrangian}
The total effective Lagrangian consists of 
\begin{eqnarray}
\lag=\mathcal{L}_{NG}
+\mathcal{L}_S + \mathcal{L}_V +\mathcal{L}_{SV},
\label{lagall}
\end{eqnarray}
where $\mathcal{L}_{NG}$ is the Lagrangian for the NG bosons,
\begin{eqnarray}
\mathcal{L}_{NG}= \frac{1}{4}{\rm Tr}[\partial^\mu \Sigma^\dag \partial_\mu \Sigma]-V(\Sigma),
\end{eqnarray}
where $V(\Sigma)$ is the potential term of $\Sigma$, and here we do not show its explicit form because this term does not affect the diquark masses after introducing a mean field of $\Sigma$.

\begin{figure}[tb!]
 \centering
  \includegraphics[width=7cm]{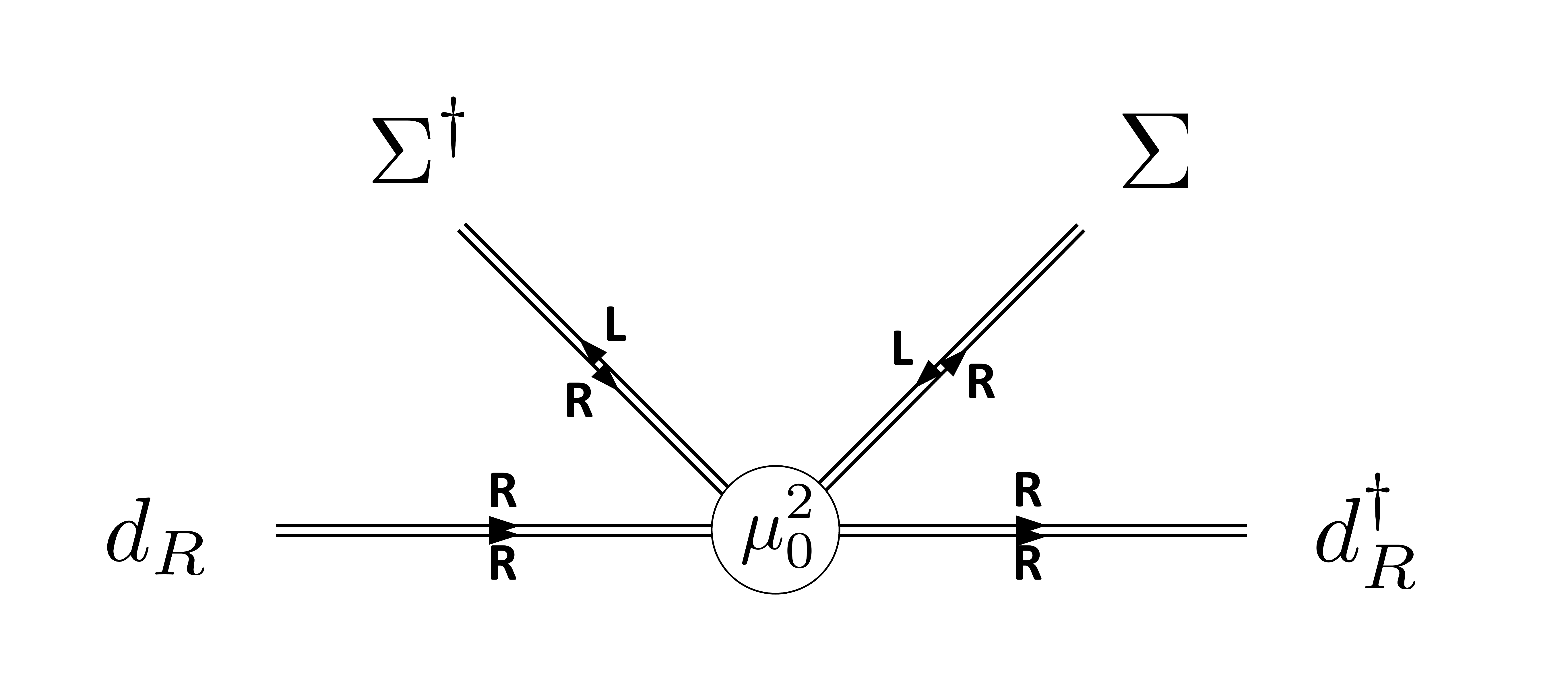}
  \caption{Quark line representation of the interaction of right-handed diquarks and mesons for the new term in the Lagrangian~(\ref{sLag_plus}).}
 \label{figure_QL}
\end{figure}

The effective Lagrangian $\mathcal{L}_S$ for the S and P diquarks is taken as
\begin{align}
\mathcal{L}_S=&\mathcal{D}_\mu d_{R,i}(\mathcal{D}^\mu d_{R,i})^\dag
		   +\mathcal{D}_\mu d_{L,i}(\mathcal{D}^\mu d_{L,i})^\dagger  \nonumber\\
		   &-m_{S0}^2(d_{R,i}d_{R,i}^\dagger+d_{L,i}d_{L,i}^\dagger) \nonumber\\
&  -\frac{m_{S1}^2}{f_\pi}(d_{R,i}\Sigma_{ij}^\dagger d_{L,j}^\dagger+d_{L,i}\Sigma_{ij}d_{R,j}^\dagger)\nonumber\\
& -\frac{m_{S2}^2}{2f_\pi^2}\epsilon_{ijk}\epsilon_{lmn} (d_{R,k}\Sigma_{li}\Sigma_{mj}d_{L,n}^\dagger +d_{L,k}\Sigma^\dag_{li}\Sigma^\dag_{m,j}d_{R,n}^\dagger)\nonumber\\
&  +\frac{\mu_0^2}{f_\pi^2}\left(d_R\{ \Sigma^\dag \Sigma-\frac{1}{3}{\rm Tr}[\Sigma^\dag \Sigma] \}d_R^\dag + \right. \nonumber\\
& \left. d_L\{ \Sigma \Sigma^\dag-\frac{1}{3}{\rm Tr}[\Sigma \Sigma^\dag] \}d_L^\dag \right). 
\label{sLag_plus}	   
\end{align}
Here, $\mathcal{D}_\mu$ in the kinetic term is the covariant derivative with the color gauge field.
The terms with $m_{S0}^2$, $m_{S1}^2$, and $m_{S2}^2$ are the same as our previous study~\cite{CETdiquark,scalar}.
The only difference is the last term with the coefficient ${\mu_0}^2$.
This term represents the eight-quark interaction described as Fig.~\ref{figure_QL}.%
\footnote{This additional term was also considered in Ref.~\cite{takada} in the effective Lagrangian for heavy baryons.
A similar term can be added to the effective Lagrangian for vector diquarks $\lag_V$.
However, it will make no change to the diquark spectrum when the flavor $SU(3)$ symmetry is broken.
So we do not introduce the new term in $\lag_{V}$.}

The Lagrangian $\mathcal{L}_V$ for the A and V diquarks are given by
\begin{align}
\mathcal{L}_V = &-\frac{1}{2}{\rm Tr}[F^{\mu\nu}F^{\dag}_{\mu\nu}] +m_{V0}^2{\rm Tr}[d^{\mu}d^{\dag}_{\mu}] \nonumber\\
 &+\frac{m_{V1}^2}{f_{\pi}^{2}}{\rm Tr}[\Sigma^{\dag}d^{\mu}\Sigma^Td^{\dag T}_{\mu}] \nonumber\\
 &+\frac{m_{V2}^2}{f_{\pi}^{2}}\left\{{\rm Tr}[\Sigma^T\Sigma^{\dag T} d_\mu^\dag d^\mu] +{\rm Tr}[\Sigma\Sigma^{\dag} d^{\mu}d^{\dag}_{\mu}]\right\},
 \label{vLag}
\end{align}
where $F^{\mu\nu} =\mathcal{D}^\mu d^{\nu} - \mathcal{D}^\nu d^{\mu}$.
This Lagrangian is the same as that taken in our previous study~\cite{Kim:2021ywp}.
Note that the Lagrangian $\mathcal{L}_{SV}$ for the coupling between S/P and A/V diquarks, shown later in Eq.~(\ref{SVLag}), also remains the same as our previous study~\cite{Kim:2022pyq}.

The four parameters, $m_{S0}^2$, $m_{S1}^2$, $m_{S2}^2$, and $\mu_0^2$ for the S/P diquarks will be determined in Sec.~\ref{Sec:parameter}.
The three parameters, $m_{V0}^2$, $m_{V1}^2$, and $m_{V2}^2$, for the A/V diquarks are given in~Ref.~\cite{Kim:2021ywp}.

We here concentrate on the S/P diquarks and their masses generated by chiral symmetry breaking.
The effective quark mass generated by spontaneous symmetry breaking is expressed as 
``$g_s\times \langle\Sigma\rangle$," 
where $g_s=2\pi/\sqrt{3}$~\cite{Delbourgo:1993dk,Scadron:2013vba} is the quark-meson coupling constant. 
We further introduce the explicit SU(3) breaking in the form of the current quark mass, $\mathcal{M}={\rm diag}(m_u, m_d, m_s)$, 
resulting in the effective quark mass, ``$\mathcal{M}+g_s\times \langle\Sigma\rangle$."
We will neglect the current light-quark masses, taking $m_u=m_d=0$.
Combining them, we have the mean field of $\Sigma$
in terms of  the flavor SU(3) breaking constant $A$
\begin{eqnarray}
&& \langle \Sigma \rangle_{SB}=f_\pi {\rm diag}(1,1,A),
\label{CSB}\\
&& A= \displaystyle\frac{f_s}{f_{\pi} }+ \frac{m_s}{g_s f_{\pi}}.
\end{eqnarray}
Note that $A$ also represents the ratio of  the constituent quark masses of $s$ and $(u,d)$ quarks.

The Lagrangian (\ref{sLag_plus}) for the S/P diquarks is invariant 
under the chiral $SU(3)_R\times SU(3)_L$ transformation given above. 
It should be noted that the diquark mass term,
$-m_{S0}^2(d_{R,i}d_{R,i}^\dagger+d_{L,i}d_{L,i}^\dagger)$, is also chiral symmetric and allows nonzero
diquark masses even when chiral symmetry breaking vanishes.
It is also important to note that the term with the coefficient $m_{S1}^2$ represents $U(1)_A$ anomaly,
which changes three right-handed quarks into three left-handed quarks and vice versa.
We have seen in the previous study that the axial anomaly by the $m_{S1}^2$ term gives significant effects on the diquark spectrum.

In the updated Lagrangian (\ref{sLag_plus}), the new term with the coefficient ${\mu_0}^2$ is both chiral $SU(3)_R\times SU(3)_L$ and $U(1)_A$ symmetric.
If chiral condensates are flavor $SU(3)$ invariant, then this term is reduced to a change of $m_{S0}^2$, 
so that no new effect is seen in the spectrum.
However, when the flavor $SU(3)$ symmetry is broken by the chiral condensates, 
the new term makes differences from the $m_{S0}^2$ term.
This is the reason why we add this new term and explore its effect in this paper.

\subsection{Diquark mass formulas}
By substituting Eq.~(\ref{CSB}) into Eq.~(\ref{sLag_plus}), we obtain the mean-field Lagrangian of the S ($0^+$)  and P ($0^-$) diquarks including the effect of the spontaneous chiral symmetry breaking. 
Then, the effective mass formulas of diquarks are
\begin{align}
&M_{ns}^2(0^+)=m_{S0}^2+\frac{1}{3}(A^2-1)\mu_0^2-m_{S1}^2-A m_{S2}^2,\label{Smass1}\\
&M_{ud}^2(0^+)=m_{S0}^2-\frac{2}{3}(A^2-1)\mu_0^2-A m_{S1}^2-m_{S2}^2,\label{Smass2}\\
&M_{ns}^2(0^-)=m_{S0}^2+\frac{1}{3}(A^2-1)\mu_0^2+m_{S1}^2+A m_{S2}^2,\label{Smass3}\\
&M_{ud}^2(0^-)=m_{S0}^2-\frac{2}{3}(A^2-1)\mu_0^2+A m_{S1}^2+m_{S2}^2,\label{Smass4}
\end{align}
where the subscript $n= u,d$ denotes the flavor of the constituent light quark.
These mass formulas are, except for the $\mu_0^2$ term, the same as those in the previous study~\cite{CETdiquark,scalar}.

In the flavor SU(3) symmetric limit, $A=1$, the effect of the additional ${\mu_0}^2$ term vanishes.
In turn, for $A>1$ and ${\mu_0}^2>0$, this term makes the strange diquarks ($us$, $ds$) 
heavier than the nonstrange diquarks ($ud$) for both the S and P diquarks.

In our previous study~\cite{CETdiquark,scalar}, we found that the P diquarks are subjected to the {\it inverse mass hierarchy} due to the $U(1)_A$ anomalous term proportional to $m_{S1}^2$.
With the new $\mu_0^2$ term,  
the mass differences between the strange and nonstrange diquarks are written as
\begin{align}
&[M_{ns}(0^+)]^2-[M_{ud}(0^+)]^2 \nonumber\\
&\ \ \ \ \ =(A-1)(m_{S1}^2-m_{S2}^2)+(A^2-1)\mu_0^2 , \label{scalar_update}\\
&[M_{ns}(0^-)]^2-[M_{ud}(0^-)]^2 \nonumber\\
&\ \ \ \ \ = - (A-1)(m_{S1}^2-m_{S2}^2) +(A^2-1)\mu_0^2. \label{pscalar_update} 
\end{align}
If ${\mu_0}^2=0$, the mass ordering of the strange and nonstrange diquarks with the negative parity is opposite to  that with the positive parity.
Therefore, by assuming that the S diquarks follow the normal ordering, $M_{ns}(0^+)>M_{ud}(0^+)$, which is necessary 
to satisfy the mass ordering of $\Lambda_Q$ and $\Xi_Q$ in ground states, the P diquarks have the inverse ordering, $M_{ns}(0^-)<M_{ud}(0^-)$.
However, since the signs of the $\mu_0^2$ terms in Eqs.~(\ref{scalar_update}) and (\ref{pscalar_update}) are the same, the inverse hierarchy may disappear 
when ${\mu_0}^2$ is large enough, that is $(A+1){\mu_0}^2>m_{S1}^2-m_{S2}^2$.
Thus, the strength of ${\mu_0}^2$ is important for determining the normal or inverse mass ordering of the P diquarks.

Furthermore, from Eqs.~(\ref{vLag}) and (\ref{CSB}), the mass formulas for the A ($1^+$) and V ($1^-$) diquarks are
\begin{align}
&M_{nn}^2 (1^+)= m_{V0}^2 + m_{V1}^2 + 2m_{V2}^2, \label{Vmass1}\\
&M_{ns}^2 (1^+)= m_{V0}^2 + A(m_{V1}^2+2m_{V2}^2), \label{Vmass2}\\
&M_{ss}^2 (1^+)= m_{V0}^2 + (2A-1)(m_{V1}^2+2m_{V2}^2), \label{Vmass3}\\
&M_{ud}^2 (1^-)= m_{V0}^2 - m_{V1}^2 + 2m_{V2}^2, \label{Vmass4}\\
&M_{ns}^2 (1^-)= m_{V0}^2 +A(-m_{V1}^2+2m_{V2}^2). \label{Vmass5}
\end{align}
These formulas are the same as those in our previous work~\cite{Kim:2021ywp}.

\begin{table*}[bt!]
  \centering
      \caption{Four parameter choices of the present work. 
We list the masses of the $\rho$-mode excited $ \Xi_Q$ and corresponding singly strange P diquarks, the parameters of effective Lagrangian (\ref{sLag_plus}), and the normal or inverse mass hierarchy of P diquarks.
    }
  \begin{tabular}{l  c  c c c c c c c c} \hline\hline 
&\multicolumn{3}{c}{Baryon/diquark masses (MeV)}  &
&\multicolumn{4}{c}{$\lag_S$ parameters (MeV$^2$)} 
& \\
\cline{2-4} \cline{6-9}
\multicolumn{1}{l }{Cases}
&$M_0 \equiv M_{\rho}(\Xi_c;1/2^-)$&$M_{\rho}(\Xi_b;1/2^-)$&$M_{ns}(0^-)$
&&$m_{S0}^2$&$m_{S1}^2$&$m_{S2}^2$&$\mu_0^2$ &
\multicolumn{1}{c}{Hierarchy} \\ \hline

Case:L&2623	&5946	&1115& &$(1062)^2$&$(762)^2$&$-(491)^2$&$-(324)^2$   & $M_{ud}(0^-) >M_{ns}(0^-)$ \\ 
Case:N~\cite{scalar}&2765	&6084	&1271&& $(1119)^2$&$(690)^2$&$-(258)^2$&$0$ & $M_{ud}(0^-) >M_{ns}(0^-)$   \\
Case:E&2890		&6207	&1406&& $(1171)^2$&$(612)^2$&$(321)^2$&$(319)^2$  & $M_{ud}(0^-) =M_{ns}(0^-)$ \\
Case:H&3279		&6589	&1819&& $(1347)^2$&$0	$&$(852)^2$&$(690)^2$  & $M_{ud}(0^-) <M_{ns}(0^-)$  \\
   \hline\hline
  \end{tabular}
  \label{Xirho_parameter}
\end{table*}

\section{Choices of parameters} \label{Sec:parameter}

To determine the four parameters, $m_{S0}^2$, $m_{S1}^2$, $m_{S2}^2$, and $\mu_0^2$, in the effective Lagrangian (\ref{sLag_plus}) for the S/P diquarks, we need four inputs, such as diquark masses.
In our previous studies~\cite{scalar} (where $\mu_0^2=0$ is fixed), we adapt the three diquark masses, $M_{ud}(0^+)=725$ MeV, $M_{ns}(0^+)=942$ MeV, and $M_{ud}(0^-)=1406$ MeV (see Model IIY in Ref.~\cite{scalar}).
In the following, we briefly summarize its procedure.
First, $M_{ud}(0^+)$ is taken from lattice QCD simulations~\cite{ChinQCD}.
Using $M_{ud}(0^+)$, a Hamiltonian for a diquark--heavy-quark two-body system is constructed so as to correctly reproduce the mass of the ground-state $\Lambda_c$, $M(\Lambda_c;1/2^+)= 2286$ MeV.
Then, $M_{ns}(0^+)$ is determined so that the same Hamiltonian reproduces the mass of the ground-state $\Xi_c$, $M(\Xi_c;1/2^+)= 2469$ MeV.
The third diquark mass, $M_{ud}(0^-)$, is related to the mass of $\rho$-mode excited $\Lambda_c(1/2^-)$ including a P diquark, 
$M_{\rho(P)}(\Lambda_c;1/2^-)$. 
Because this state is not experimentally observed so far, 
we take the value $M_{\rho(P)}(\Lambda_c;1/2^-)=2890$ MeV, obtained from a three-body calculation of a constituent quark model in Ref.~\cite{Yoshida}.

The fourth input is the strange P-diquark mass, $M_{ns}(0^-)$.\footnote{In our previous study~\cite{scalar}, $M_{ns}(0^-)$ is obtained from the diquak mass formulas after determining the three parameters, ($M_{ud}(0^+)$, $M_{ns}(0^+)$, $M_{ud}(0^-)$), or ($m_{S0}^2$, $m_{S1}^2$, $m_{S2}^2$).
In this sense, $M_{ns}(0^-)$ is regarded as a prediction.
On the other hand, in the current work with four parameters, $M_{ns}(0^-)$ is treated as an input parameter.}
To determine this value, we assume the mass of the $\rho$-mode excited state of $\Xi_c$ within the following range:
\begin{align}
2623~{\rm MeV} < M_{\rho(P)}(\Xi_c;1/2^-)< 3279~{\rm MeV}.
\label{M0param}
\end{align}
Hereafter, we denote $M_0 \equiv M_{\rho(P)}(\Xi_c;1/2^-)$, and 
$M_0$ is treated as the input parameter changing within the range of Eq.~(\ref{M0param}).

The lower limit of Eq.~(\ref{M0param}) is given so that the strange A diquark is lighter than the strange P diquark as $M_{ns}(0^-) > M_{ns}(1^+)=1115.30$ MeV.
Then, the lower limit of $M_0$ is given by 2623 MeV, same as the spin-averaged mass of the ground-state $\Xi'_c$ baryons, $M(\Xi'_c;1/2^+,3/2^+)= 2622.903$ MeV.
On the other hand, the upper limit of Eq.~(\ref{M0param}) is chosen so that $m_{S1}^2\ge 0$,\footnote{If $m_{S1}^2< 0$, the nonstrange S/P diquark masses become $M_{ud}(0^-) < M_{ud}(0^+)$ in the chiral-symmetry restored phase.
To avoid this situation, we constrain $m_{S1}^2\ge 0$.} which determines the upper limit of $M_0$ as 3279 MeV.
Then, the mass of the P diquark is $M_{ns}(0^-)<1818.658$ MeV.
Thus, the mass range of the strange P diquark, corresponding to the range~(\ref{M0param}), is given as
\begin{align}
1115~{\rm MeV} <M_{ns}(0^-)< 1819~ {\rm MeV}.
\end{align}
We also see that the same parameter set allows the $\rho$-mode $\Xi_b$ baryon to have the mass in the range of
 5946 MeV ${< M_{\rho(P)}(\Xi_b;1/2^-)<}$ 6589 MeV.

In this work, we do not further restrict our parameters, but instead
we will focus on the following four special cases:
\begin{align}
&\mathrm{Case:L}, \ M_0=2623 \mathrm{MeV} &&\mathrm{[for~} M_{ns}(0^-)=M_{ns}(1^+)], \nonumber\\
&\mathrm{Case:N}, \ M_0=2765 \mathrm{MeV} &&\mathrm{[for~} \mu_0^2=0], \nonumber\\
&\mathrm{Case:E}, \ M_0=2890 \mathrm{MeV} &&\mathrm{[for~} M_{ud}(0^-)=M_{ns}(0^-)], \nonumber\\
&\mathrm{Case:H}, \ M_0=3279 \mathrm{MeV} &&\mathrm{[for~} m_{S1}^2=0],\nonumber
\end{align}
where ``L," ``N," ``E," and ``H" mean the initials of ``Lower," ``No," ``Equal," and ``Higher," respectively.

The baryon/diquark masses and the corresponding Lagrangian parameters are summarized in Table \ref{Xirho_parameter}.
For Cases:L and N, the inverse hierarchy $M_{ud}(0^-)>M_{ns}(0^-)$ is realized,
while in Case:E, they are equal, $M_{ud}(0^-)=M_{ns}(0^-)=1406$ MeV.
In Case:H, we have the normal ordering $M_{ud}(0^-)<M_{ns}(0^-)$.

\begin{figure*}[htp]
  \includegraphics[width=10cm]{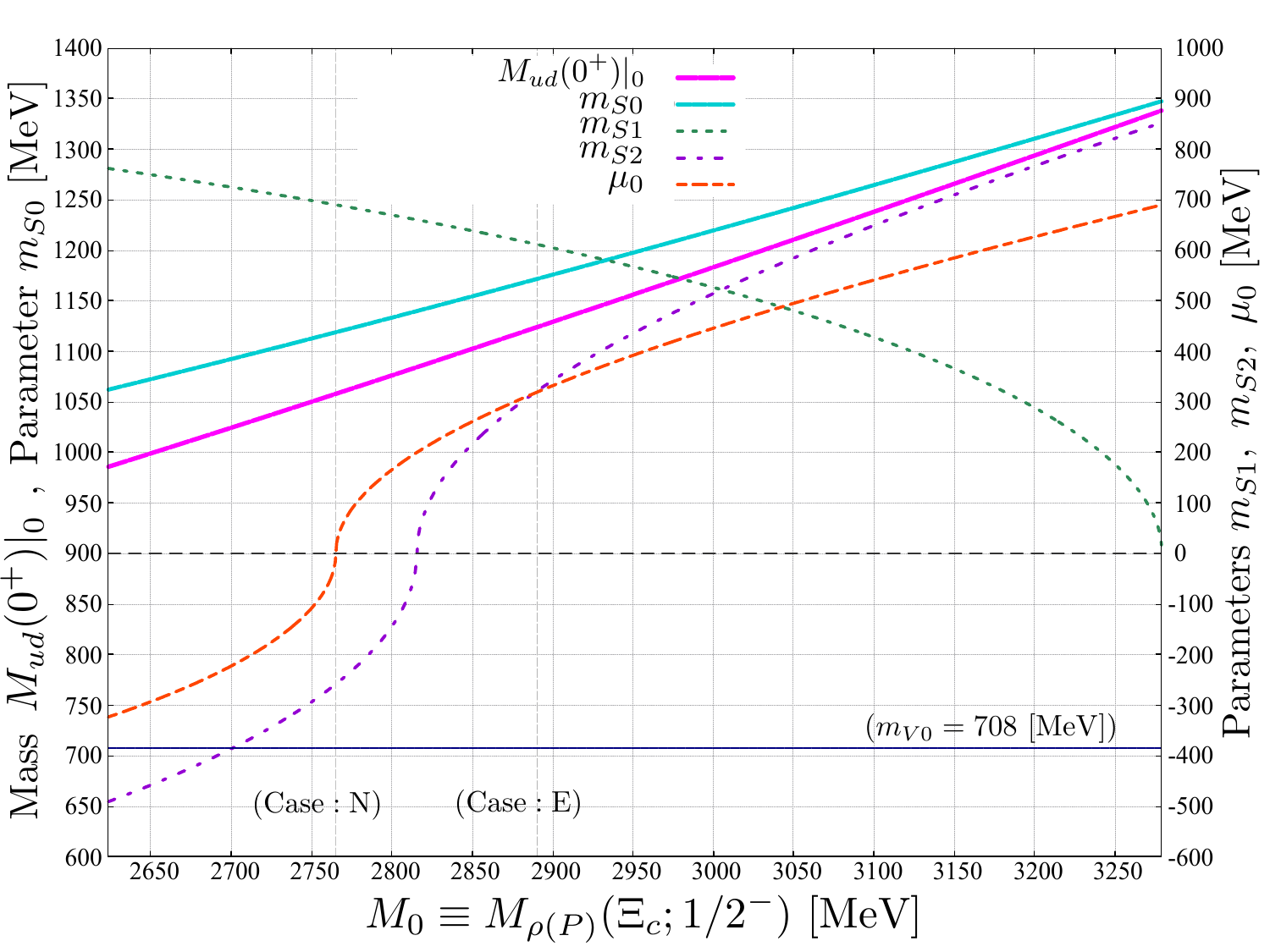}
  \caption{Parameters in new effective Lagrangian~(\ref{sLag_plus}) of S/P diquarks and the mass of nonstrange S diquark in the chiral symmetry restored phase, Eq. (\ref{x=0scalar}), as functions of mass $M_0$ in the range of Eq. (\ref{M0param}).
Two vertical lines, $M_0=2765$ and 2890 MeV, stand for Case:N and Case:E, respectively.
The solid line parallel to the horizontal line denotes the parameter $m_{V0}=708$ MeV showing the chiral invariant mass of nonstrange A diquark.
Note that $m_{S2}$ and $\mu_0$ are plotted as ${\rm sgn}(m^2)\times|m|$.
}
 \label{figure_Lagadd}
\end{figure*} 

Figure~\ref{figure_Lagadd} shows the values of the Lagrangian parameters as functions of $M_0$,
where the solid and dashed lines obey the left and right side of the vertical axis, respectively. 
When $M_0$ increases, $m_{S0}$, $m_{S2}$, and $\mu_0$ increase while $m_{S1}$ decreases. 
The parameters, $\mu_0$ and $m_{S2}$, vanish at $M_0=2765$ MeV (Case:N) and $M_0=2815$ MeV, respectively. 

It is interesting to see the behavior of the nonstrange S-diquark mass in the chiral-symmetry restored phase
\begin{align}
M_{ud}^2(0^+)|_{0}= m_{S0}^2 -\frac{2}{3}\left(\frac{m_s}{g_s f_\pi}\right)^2\mu_0^2 -\frac{m_s}{g_s f_\pi}m_{S1}^2, 
\label{x=0scalar}
\end{align}
where ``$...|_{0}$" denotes the limit of chiral restoration [see Eq.~(\ref{newSmassx}) in the next section, for details].
This behavior is also plotted as the magenta solid line in Fig.~\ref{figure_Lagadd}.
We can see $M_{ud}(0^+)|_{0}$ is not only given by $m_{S0}$ but also contains contributions coming from the the $m_{S1}$ and $\mu_0$ terms of the Lagrangian.

Because the parameter $m_{S1}^2$ is positive and larger than $\mu_0^2$ for $\mu_0^2<0$, $M_{ud}(0^+)|_{0}$ is smaller than $m_{S0}$ for all mass range of $M_0$ in this work. With the increase of $M_0$, the effect of parameter $m_{S1}$ weakens and thus the difference between its mass $M_{ud}(0^+)|_{0}$ and parameter $m_{S0}$ shrinks. 
On the other hand, $M_{ud}(0^+)|_{0}$ is always larger than the mass of A diquark in the same chiral symmetric limit, $m_{V0}=708$ MeV. 

\section{Diquarks under chiral restoration} \label{Sec:restoration}

Our main interest in studying chiral effective theory of diquarks is 
how the masses of diquarks change in hot and/or dense matter. 
It may appear in the singly-heavy-baryon spectrum in hot/dense matter.
For instance, the change of the mass difference between the A and S diquarks will directly affect
the decay widths of $\Sigma_Q \to \Lambda_Q+\pi$.
The widths of the $\rho$-mode excited states of baryons may also change.

In this work, we do not calculate the temperature or density dependence within the present model, but instead we consider the change of the chiral condensate and examine its effects on diquark spectra.
We assume that the chiral condensates in Eq.~(\ref{CSB}) are scaled by a parameter $x$ ($x=1\to 0$), while the strange quark mass is independent of $x$, as%
\begin{align}
\langle \Sigma \rangle(x)= {\rm diag}(xf_\pi,xf_\pi,\frac{m_s}{g_s}+xf_s).
\end{align}
Under this assumption, the $x$-dependent mass formulas of the nonstrange ($ud$) S/P diquarks are written as
\begin{align}
M_{ud}^2(0^+)=&m_{S0}^2-\frac{2}{3}\left\{\left(x\frac{f_s}{f_\pi}+\frac{m_s}{g_s f_\pi}\right)^2-x^2\right\}\mu_0^2 \nonumber\\
&-\left(x\frac{f_s}{f_\pi}+\frac{m_s}{g_s f_\pi} \right)m_{S1}^2-x^2m_{S2}^2,~~ \label{newSmassx}\\
M_{ud}^2(0^-)=& m_{S0}^2-\frac{2}{3}\left\{\left(x\frac{f_s}{f_\pi}+\frac{m_s}{g_s f_\pi}\right)^2-x^2\right\}\mu_0^2\nonumber\\
&+\left(x\frac{f_s}{f_\pi}+\frac{m_s}{g_s f_\pi} \right)m_{S1}^2+x^2m_{S2}^2,~~\label{newPmassx}
\end{align}
while those of the nonstrange A/V diquarks are~\cite{Kim:2021ywp}
\begin{align}
M_{nn}^2(1^+)=& m_{V0}^2 + x^2(m_{V1}^2+2m_{V2}^2), \label{Amassx}\\
M_{ud}^2(1^-)=& m_{V0}^2 + x^2(-m_{V1}^2+2m_{V2}^2). \label{Vmassx}
\end{align}

Here, we comment on an error in the $x$-dependent mass formulas used in our previous papers~\cite{Kim:2021ywp,Kim:2022mpa,Kim:2022pyq}, where the $m_{S0}^2$, $m_{S1}^2$, and $m_{S2}^2$ terms are included but the $\mu_0^2$ term is not.
The coefficients of $m_{S1}^2$ in Eq.~(38) in Ref.~\cite{Kim:2021ywp},
Eqs.~(28) and (29) in Ref.~\cite{Kim:2022mpa}, and 
Eq.~(50) in Ref.~\cite{Kim:2022pyq}
are incorrect. 
The correct coefficient is given as Eqs.~(\ref{newSmassx}) and (\ref{newPmassx}).

\begin{figure}[tb!]
\centering
  \includegraphics[clip,width=1.0\columnwidth]{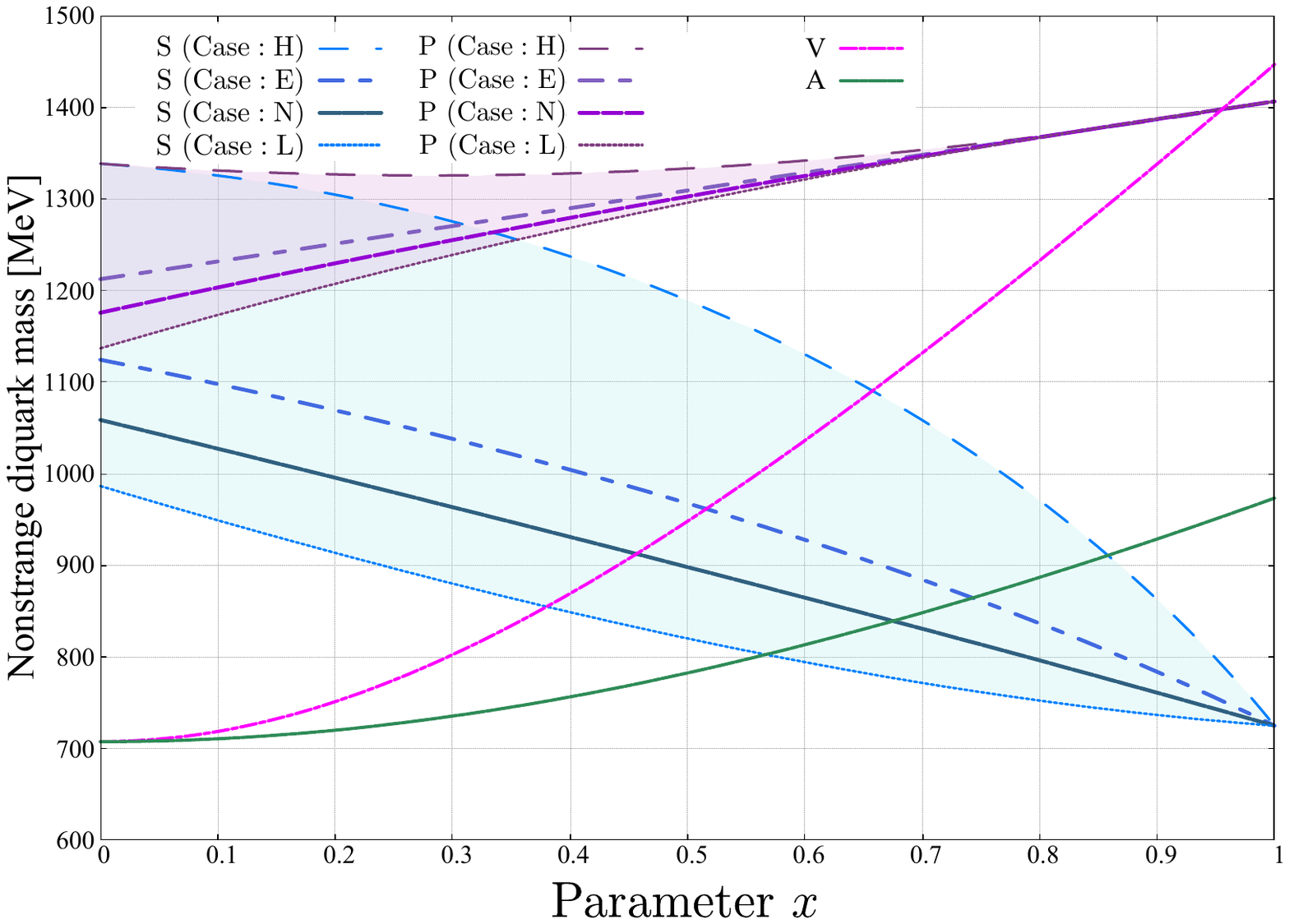}
  \caption{Dependence of nonstrange ($ud$, $uu$, or $dd$) diquark masses on the chiral symmetry breaking parameter $x$,
obtained from Eqs.~(\ref{newSmassx})--(\ref{Vmassx}) and the parameter choices, Case:L, Case:N, Case:E, and Case:H.
The cyan and purple shaded regions designate the possible ranges of the S- and P-diquark masses, respectively.
Also see Fig.~1 in Ref.~\cite{Kim:2021ywp} as the previous study.
  }
 \label{figure_Lagadd3}
\end{figure}

The $x$ dependences of the masses of the nonstrange S/P and A/V diquarks 
are plotted in Fig.~\ref{figure_Lagadd3} for the parameter choices, Case:L, Case:N, Case:E, and Case:H.\footnote{Note that the input mass $M_0$ increases in this order.
One sees that both the S/P-diquark masses increase as $M_0$ increases, since the parameter $m_{S0}$ increases as shown in Fig. \ref{figure_Lagadd}.}
As $x$ decreases, the S-diquark mass increases: the S diquark gets heavier under a chiral restoration.
On the other hand, the P-diquark mass decreases, and hence the mass difference between the S and P diquarks shrinks, which reflects the chiral-partner structure of S/P diquarks.
It should be noted that the S and P diquarks are not degenerate even at $x=0$, because of the $U(1)_A$ anomaly effect represented by the parameter $m_{S1} \neq 0$.
However, only for Case:H, this mass difference at $x=0$ vanishes due to $m_{S1}=0$.

For the A/V diquarks, both the masses decrease as $x$ decreases, indicating that the A/V diquarks become lighter
in chiral restored medium.
At $x=0$, the masses of A and V diquarks are completely degenerate, which is attributed to the chiral-partner structure of A/V diquarks.

In addition, we comment on the {\it lowest-state inversion}~\cite{Kim:2021ywp}, where we compare the masses of nonstrange A and S diquarks.
The S diquark is lighter than the A diquark in the ordinary vacuum at $x=1$, whereas the A diquark becomes lighter than the S diquark in the chiral-symmetry restored phase at $x=0$.
From Fig.~\ref{figure_Lagadd3}, we can see a crossing point where the lines of the A- and S-diquark masses cross with each other. 
The value of $x$ of this crossing point is estimated to be ${x_{AS}= 0.6747}$ in Case:N.\footnote{In our previous study~\cite{Kim:2021ywp}, we obtained $x_{AS} \sim 0.6$ using the same parameters (see Fig.~1 in Ref.~\cite{Kim:2021ywp}), but this result is influenced by the error in the mass formulas.
The correct result is Case:N in Fig.~\ref{figure_Lagadd3}.}
Using the results in Case:L and Case:H, the possible range of $x_{AS}$ is estimated to be $0.5661< x_{AS} <0.8580$.

\begin{figure}[tb!]
\centering
  \includegraphics[clip,width=1.0\columnwidth]{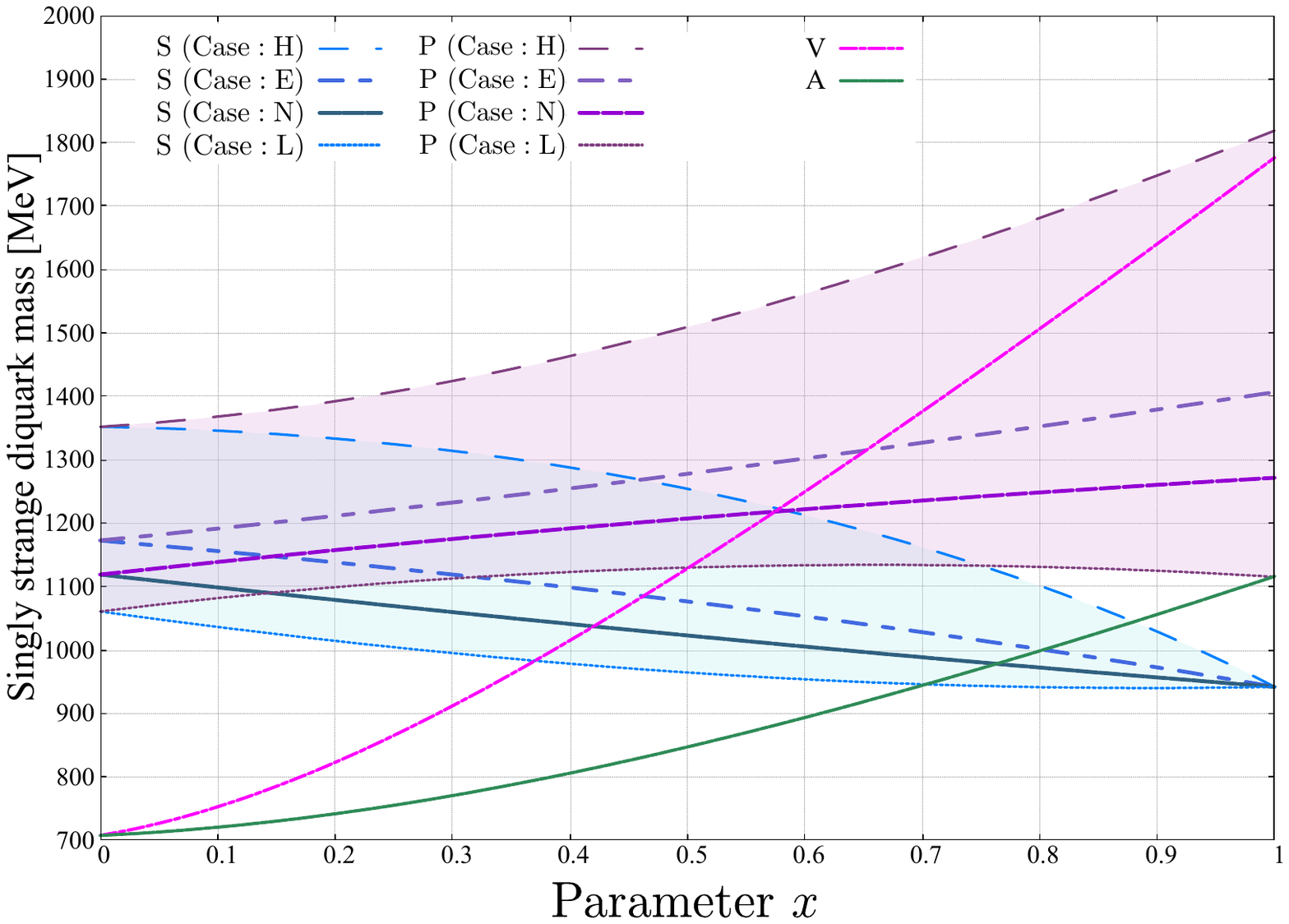}
  \caption{Dependence of singly strange ($us$ or $ds$) diquark masses on the chiral symmetry breaking parameter $x$,
obtained from Eqs.~(\ref{Ssnmassx})--(\ref{Vsnmassx}) and the parameter choices, Case:L, Case:N, Case:E, and Case:H.
The cyan and purple shaded regions designate the possible ranges of the S- and P-diquark masses, respectively.
  }
 \label{figure_Lagadd3s}
\end{figure}

In the case of singly strange ($us$ or $ds$) S and P diquarks, the $x$-dependent mass formulas are given by
\begin{align}
M_{ns}^2(0^+)=&m_{S0}^2+\frac{1}{3}\left\{\left(x\frac{f_s}{f_\pi}+\frac{m_s}{g_s f_\pi}\right)^2 -x^2\right\}\mu_0^2 \nonumber\\
&-xm_{S1}^2-x\left(x\frac{f_s}{f_\pi}+\frac{m_s}{g_s f_\pi}\right)m_{S2}^2 ,\label{Ssnmassx}\\
M_{ns}^2(0^-)=& m_{S0}^2+\frac{1}{3}\left\{\left(x\frac{f_s}{f_\pi}+\frac{m_s}{g_s f_\pi}\right)^2-x^2\right\}\mu_0^2  \nonumber\\
&+xm_{S1}^2+x\left(x\frac{f_s}{f_\pi}+\frac{m_s}{g_s f_\pi}\right)m_{S2}^2 ,\label{Psnmassx}
\end{align}
while those of singly strange A and V diquarks are\footnote{Note that the second order of $\epsilon(x)=x\frac{f_s}{f_\pi}+\frac{m_s}{g_s f_\pi}-x$ is neglected in Eqs. (\ref{Asnmassx}) and (\ref{Vsnmassx}).}
\begin{align}
M_{ns}^2(1^+)=& m_{V0}^2 + x^2(m_{V1}^2+2m_{V2}^2) \nonumber\\
&+x\left(x\frac{f_s}{f_\pi}+\frac{m_s}{g_s f_\pi}-x\right)(m_{V1}^2+2m_{V2}^2) , \label{Asnmassx}\\
M_{ns}^2(1^-)=& m_{V0}^2 + x^2(-m_{V1}^2+2m_{V2}^2)\nonumber\\
&+x\left(x\frac{f_s}{f_\pi}+\frac{m_s}{g_s f_\pi}-x\right)(-m_{V1}^2+2m_{V2}^2) . \label{Vsnmassx}
\end{align}
The $x$ dependences from these formulas are plotted in Fig. \ref{figure_Lagadd3s}.
A difference from nonstrange diquarks is that, when chiral symmetry is restored ($x=0$), the masses of singly strange S and P diquarks are degenerate.
This is because the coefficient of $m_{S1}^2$ term in Eqs.~(\ref{Ssnmassx}) and (\ref{Psnmassx}) is only $x$, which is in contrast to the case of nonstrange diquarks in Eqs.~(\ref{newSmassx}) and (\ref{newPmassx}).
It is also noted that all the vector diquarks, independently from the flavor and parity, are degenerate in the chiral restoration limit.

\section{Heavy-baryon spectra and chiral restoration} \label{Sec:spectra}
\begin{figure*}[t!]
\centering
  \includegraphics[width=12cm]{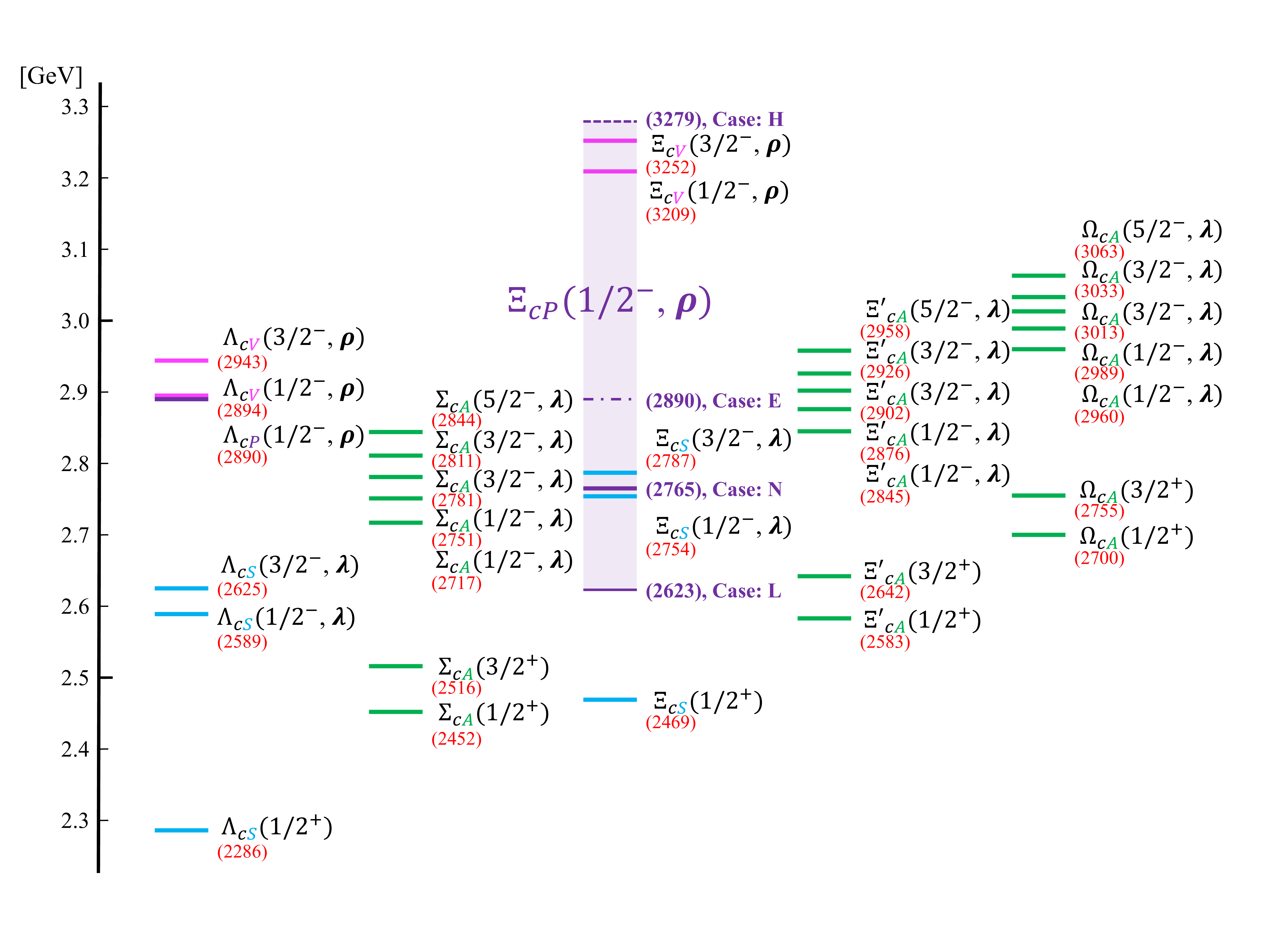}
  \caption{Mass spectra of singly charmed baryons (see Fig.~2 in Ref.~\cite{Kim:2021ywp} as the previous study).
The colors of lines show the types of constituent diquarks as S (cyan), P (purple), V (magenta), and A (green).
For the negative-parity states, they are classified into the $\rho$ mode and the $\lambda$ mode with the symbols $\rho$ and $\lambda$, respectively. The mass of $\rho$-mode excited $\Xi_c$ baryon including a P diquark, $\Xi_{cP}(1/2^-,\bm{\rho})$, is treated as a parameter varying within the shaded range.
}
 \label{figure_Cqqspectrum}
\end{figure*}

\begin{figure*}[t!]
\centering
  \includegraphics[width=12cm]{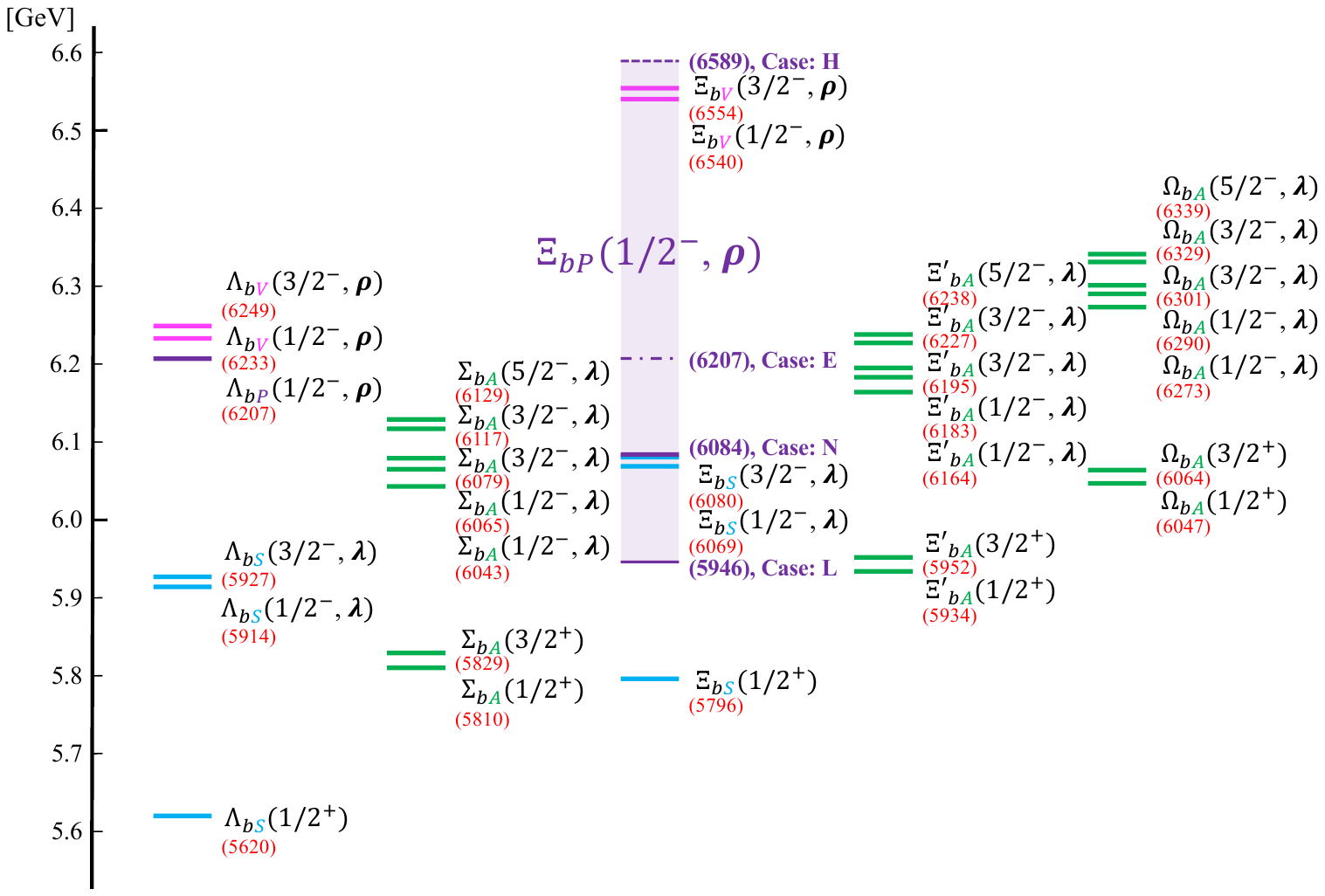}
  \caption{Mass spectra of singly bottom baryons (see Fig.~4 in Ref.~\cite{Kim:2021ywp} as the previous study).
The meanings of colors and symbols are the same as those in Fig.~\ref{figure_Cqqspectrum}.
}
 \label{figure_Bqqspectrum}
\end{figure*}

The mass spectra of singly heavy baryons can be calculated using the heavy-quark--diquark potential model.
The spectra within our model were already shown in our previous paper~\cite{Kim:2021ywp}, but here we discuss the update of the spectra.
In Figs.~\ref{figure_Cqqspectrum} and \ref{figure_Bqqspectrum}, the spectra of charmed and bottom baryons are shown, respectively.
In this calculation, the masses of $\Xi_{cP}(1/2^-,\bm{\rho})$ and $\Xi_{bP}(1/2^-,\bm{\rho})$, shown as the purple lines, are the input parameter, which varies from Case:L to Case:H.
Note that the other $\Xi_Q$'s, such as the positive-parity states and $\lambda$-mode excited states, are not affected by the parameter choice.

\begin{table}[b!]
\begin{threeparttable}
  \centering
    \caption{Masses of singly charmed baryons in the mass spectra (Fig.~\ref{figure_Cqqspectrum}) compared with the experimental data. For the baryon mass, (cal.) and (exp.) represent our calculated results and experimental data in PDG~\cite{ParticleDataGroup:2024cfk}, respectively.
Note that the mass of $\rho$-mode excited $\Lambda_c$ baryon with the constituent P diquark, 2890 MeV, is quoted from the three-quark model~\cite{Yoshida}.}
  \begin{tabular}{ c  l  c  c  c } \hline\hline
        \multicolumn{1}{ c }{}
      &\multicolumn{1}{ c }{}  
      &\multicolumn{1}{ c }{Constituent}  
      & \multicolumn{2}{c }{Mass (MeV)} \\
\cline{4-5}
        \multicolumn{1}{ c }{Baryon ($J^P$)}
      &\multicolumn{1}{ c }{State}
      &\multicolumn{1}{ c }{diquark}  
      &\multicolumn{1}{ c }{(cal.)}  
      & \multicolumn{1}{c }{(exp.)} 
      
      \\ \hline  
    $\Lambda_c$ ($1/2^+$)   	&Ground 			&S &2286\tnote{a}  	& 2286.46 
    \\
    $\Lambda_c$ ($1/2^-$)   	&$\lambda$ mode	&S &2589  	& (2592.25) 
    \\
    $\Lambda_c$ ($3/2^-$)   	&$\lambda$ mode	&S &2625  	& (2628.00) 
    \\
    $\Lambda_c$ ($1/2^-$)   	&$\rho$ mode		&P &2890  	& $\ldots$ 
    \\
    $\Lambda_c$ ($1/2^-$)   	&$\rho$ mode		&V &2894  	& $\ldots$ 
    \\
    $\Lambda_c$ ($3/2^-$)   	&$\rho$ mode		&V &2943  	& (2939.60) 
    \\
\hline
    $\Sigma_c$ ($1/2^+$)    	&Ground			&A &2452  	& 2453.46
    \\
    $\Sigma_c$ ($3/2^+$)   	&Ground			&A &2516  	& 2518.10
    \\
    $\Sigma_c$ ($1/2^-$)    	&$\lambda$ mode	&A &2717  	& $\ldots$
    \\
    $\Sigma_c$ ($1/2^-$)    	&$\lambda$ mode	&A &2751  	& $\ldots$
    \\
    $\Sigma_c$ ($3/2^-$)   		&$\lambda$ mode	&A &2781  	& $\ldots$
    \\
    $\Sigma_c$ ($3/2^-$)   		&$\lambda$ mode	&A &2811  	& $\ldots$
    \\
    $\Sigma_c$ ($5/2^-$)   		&$\lambda$ mode	&A &2844  	& $\ldots$
    \\   
 \hline   
    $\Xi_c$ ($1/2^+$)   		&Ground 			&S &2469\tnote{a}  	& 2469.08 
    \\
    $\Xi_c$ ($1/2^-$)   		&$\lambda$ mode	&S &2754  	& (2792.90) 
    \\
    $\Xi_c$ ($3/2^-$)   		&$\lambda$ mode	&S &2787  	& (2818.15) 
    \\
    $\Xi_c$ ($1/2^-$)   		&$\rho$ mode		&P &{\bf 2623--3279}  	& $\ldots$ 
    \\
    $\Xi_c$ ($1/2^-$)   		&$\rho$ mode		&V &3209  	& $\ldots$ 
    \\
    $\Xi_c$ ($3/2^-$)   		&$\rho$ mode		&V &3252  	& $\ldots$ 
    \\
\hline
    $\Xi_c'$ ($1/2^+$)    		&Ground			&A &2583  	& 2578.45
    \\
    $\Xi_c'$ ($3/2^+$)   		&Ground			&A &2642  	& 2645.63
    \\
    $\Xi_c'$ ($1/2^-$)    		&$\lambda$ mode	&A &2845  	& $\ldots$
    \\
    $\Xi_c'$ ($1/2^-$)    		&$\lambda$ mode	&A &2876  	& $\ldots$
    \\
    $\Xi_c'$ ($3/2^-$)   		&$\lambda$ mode	&A &2902  	& $\ldots$
    \\
    $\Xi_c'$ ($3/2^-$)   		&$\lambda$ mode	&A &2926  	& $\ldots$
    \\
    $\Xi_c'$ ($5/2^-$)   		&$\lambda$ mode	&A &2958  	& $\ldots$
    \\   
 \hline   
    $\Omega_c$ ($1/2^+$)    	&Ground			&A &2700  	& 2695.20
    \\
    $\Omega_c$ ($3/2^+$)   	&Ground			&A &2755  	& 2765.90
    \\
    $\Omega_c$ ($1/2^-$)    	&$\lambda$ mode	&A &2960  	& $\ldots$
    \\
    $\Omega_c$ ($1/2^-$)    	&$\lambda$ mode	&A &2989  	& $\ldots$
    \\
    $\Omega_c$ ($3/2^-$)   	&$\lambda$ mode	&A &3013  	& $\ldots$
    \\
    $\Omega_c$ ($3/2^-$)   	&$\lambda$ mode	&A &3033  	& $\ldots$
    \\
    $\Omega_c$ ($5/2^-$)   	&$\lambda$ mode	&A &3063  	& $\ldots$
    \\ \hline\hline
  \end{tabular}
\begin{tablenotes}[para,flushleft,online,normal] 
\raggedright
\item[a] Denotes the input values from PDG. 
\end{tablenotes}
  \label{Cmass_exp}
\end{threeparttable}
\end{table}

In Tables \ref{Cmass_exp} and \ref{Bmass_exp}, we summarize the numerical values of our predictions and the known experimental data from the Particle Data Group (PDG)~\cite{ParticleDataGroup:2024cfk} in 2024.
Our predictions except for $\Xi_{cP}(1/2^-,\bm{\rho})$ and $\Xi_{bP}(1/2^-,\bm{\rho})$ are the same as those in our previous paper~\cite{Kim:2021ywp} in 2021, but the experimental values are updated to the most recent ones.\footnote{In 2023, the LHCb experiment~\cite{LHCb:2023zpu} reported two new states of $\Xi_b^0$ with the masses of $6087.2$ and $6095.3$ MeV, respectively.
In Table~\ref{Bmass_exp}, in order to compare the isospin average of $\Xi_b (3/2^-)$ (i.e., the average of $\Xi_b^0$ and $\Xi_b^-$), we choose $6095.3$ MeV for $\Xi_b^0 (3/2^-)$ and $6099.8$ MeV for $\Xi_b^- (3/2^-)$.}

\begin{table}[t!]
\begin{threeparttable}
  \centering
    \caption{Masses of singly bottom baryons in the mass spectra (Fig.~\ref{figure_Bqqspectrum}) compared with the experimental data. For the baryon mass, (cal.) and (exp.) represent our calculated results and experimental data in PDG~\cite{ParticleDataGroup:2024cfk}, respectively.}
  \begin{tabular}{ c  l  c  c  c } \hline\hline
        \multicolumn{1}{ c }{}
      &\multicolumn{1}{ c }{}  
      &\multicolumn{1}{ c }{Constituent}  
      & \multicolumn{2}{c }{Mass (MeV)} \\
\cline{4-5}
        \multicolumn{1}{ c }{Baryon ($J^P$)}
      &\multicolumn{1}{ c }{State}
      &\multicolumn{1}{ c }{diquark}  
      &\multicolumn{1}{ c }{(cal.)}  
      & \multicolumn{1}{c }{(exp.)} 
            
      \\ \hline  
    $\Lambda_b$ ($1/2^+$)   	&Ground 			&S &5620\tnote{a}  	& 5619.60 
    \\
    $\Lambda_b$ ($1/2^-$)   	&$\lambda$ mode	&S &5914  	& (5912.19) 
    \\
    $\Lambda_b$ ($3/2^-$)   	&$\lambda$ mode	&S &5927  	& (5920.09) 
    \\
    $\Lambda_b$ ($1/2^-$)   	&$\rho$ mode		&P &6207  	& $\ldots$ 
    \\
    $\Lambda_b$ ($1/2^-$)   	&$\rho$ mode		&V &6233  	& $\ldots$ 
    \\
    $\Lambda_b$ ($3/2^-$)   	&$\rho$ mode		&V &6249  	& $\ldots$ 
    \\
\hline
    $\Sigma_b$ ($1/2^+$)    	&Ground			&A &5810  	& 5813.10
    \\
    $\Sigma_b$ ($3/2^+$)   	&Ground			&A &5829  	& 5832.53
    \\
    $\Sigma_b$ ($1/2^-$)    	&$\lambda$ mode	&A &6043  	& $\ldots$
    \\
    $\Sigma_b$ ($1/2^-$)    	&$\lambda$ mode	&A &6065  	& $\ldots$
    \\
    $\Sigma_b$ ($3/2^-$)   		&$\lambda$ mode	&A &6079  	& $\ldots$
    \\
    $\Sigma_b$ ($3/2^-$)   		&$\lambda$ mode	&A &6117  	& $\ldots$
    \\
    $\Sigma_b$ ($5/2^-$)   		&$\lambda$ mode	&A &6129  	& $\ldots$
    \\   
 \hline   
    $\Xi_b$ ($1/2^+$)   		&Ground 			&S &5796  	& 5794.45 
    \\
    $\Xi_b$ ($1/2^-$)   		&$\lambda$ mode	&S &6069  	& $\ldots$ 
    \\
    $\Xi_b$ ($3/2^-$)   		&$\lambda$ mode	&S &6080  	& (6097.55) 
    \\
    $\Xi_b$ ($1/2^-$)   		&$\rho$ mode		&P &{\bf 5946--6589}  	& $\ldots$ 
    \\
    $\Xi_b$ ($1/2^-$)   		&$\rho$ mode		&V &6540  	& $\ldots$ 
    \\
    $\Xi_b$ ($3/2^-$)   		&$\rho$ mode		&V &6554  	& $\ldots$ 
    \\
\hline
    $\Xi_b'$ ($1/2^+$)    		&Ground			&A &5934  	& 5935.10
    \\
    $\Xi_b'$ ($3/2^+$)   		&Ground			&A &5952  	& 5954.00
    \\
    $\Xi_b'$ ($1/2^-$)    		&$\lambda$ mode	&A &6164  	& $\ldots$
    \\
    $\Xi_b'$ ($1/2^-$)    		&$\lambda$ mode	&A &6183  	& $\ldots$
    \\
    $\Xi_b'$ ($3/2^-$)   		&$\lambda$ mode	&A &6195  	& $\ldots$
    \\
    $\Xi_b'$ ($3/2^-$)   		&$\lambda$ mode	&A &6227  	& $\ldots$
    \\
    $\Xi_b'$ ($5/2^-$)   		&$\lambda$ mode	&A &6238  	& $\ldots$
    \\   
 \hline   
    $\Omega_b$ ($1/2^+$)    	&Ground			&A &6047  	& 6045.80
    \\
    $\Omega_b$ ($3/2^+$)   	&Ground			&A &6064  	& $\ldots$
    \\
    $\Omega_b$ ($1/2^-$)    	&$\lambda$ mode	&A &6273  	& $\ldots$
    \\
    $\Omega_b$ ($1/2^-$)    	&$\lambda$ mode	&A &6290  	& $\ldots$
    \\
    $\Omega_b$ ($3/2^-$)   	&$\lambda$ mode	&A &6301  	& $\ldots$
    \\
    $\Omega_b$ ($3/2^-$)   	&$\lambda$ mode	&A &6329  	& $\ldots$
    \\
    $\Omega_b$ ($5/2^-$)   	&$\lambda$ mode	&A &6339  	& $\ldots$
    \\ \hline\hline
  \end{tabular}
\begin{tablenotes}[para,flushleft,online,normal]
\raggedright
\item[a] Denotes the input values from PDG. 
\end{tablenotes}
  \label{Bmass_exp}
\end{threeparttable}
\end{table}

\begin{figure*}[tbp!]
\centering
  \includegraphics[clip,width=1.0\columnwidth]{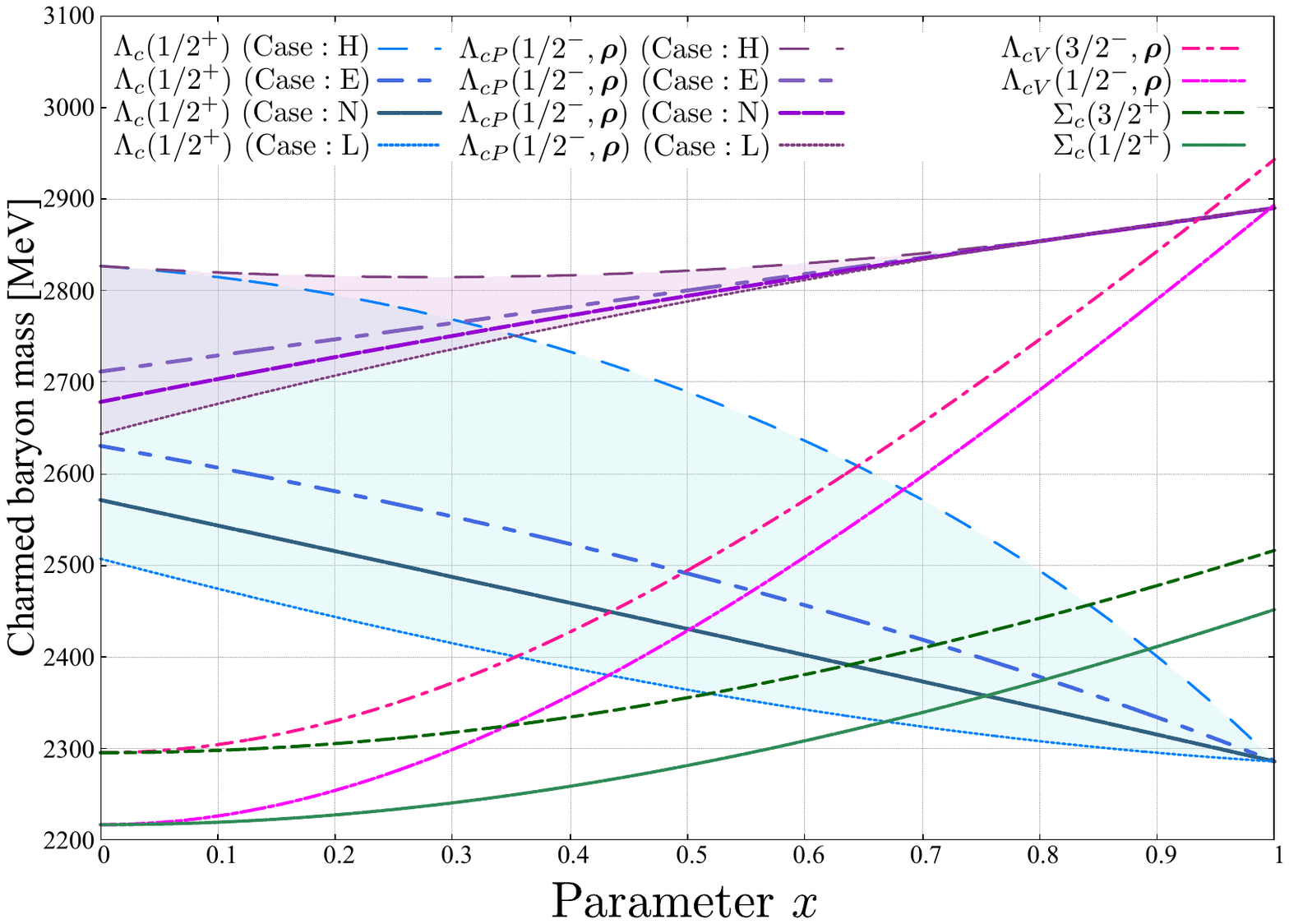}
  \includegraphics[clip,width=1.0\columnwidth]{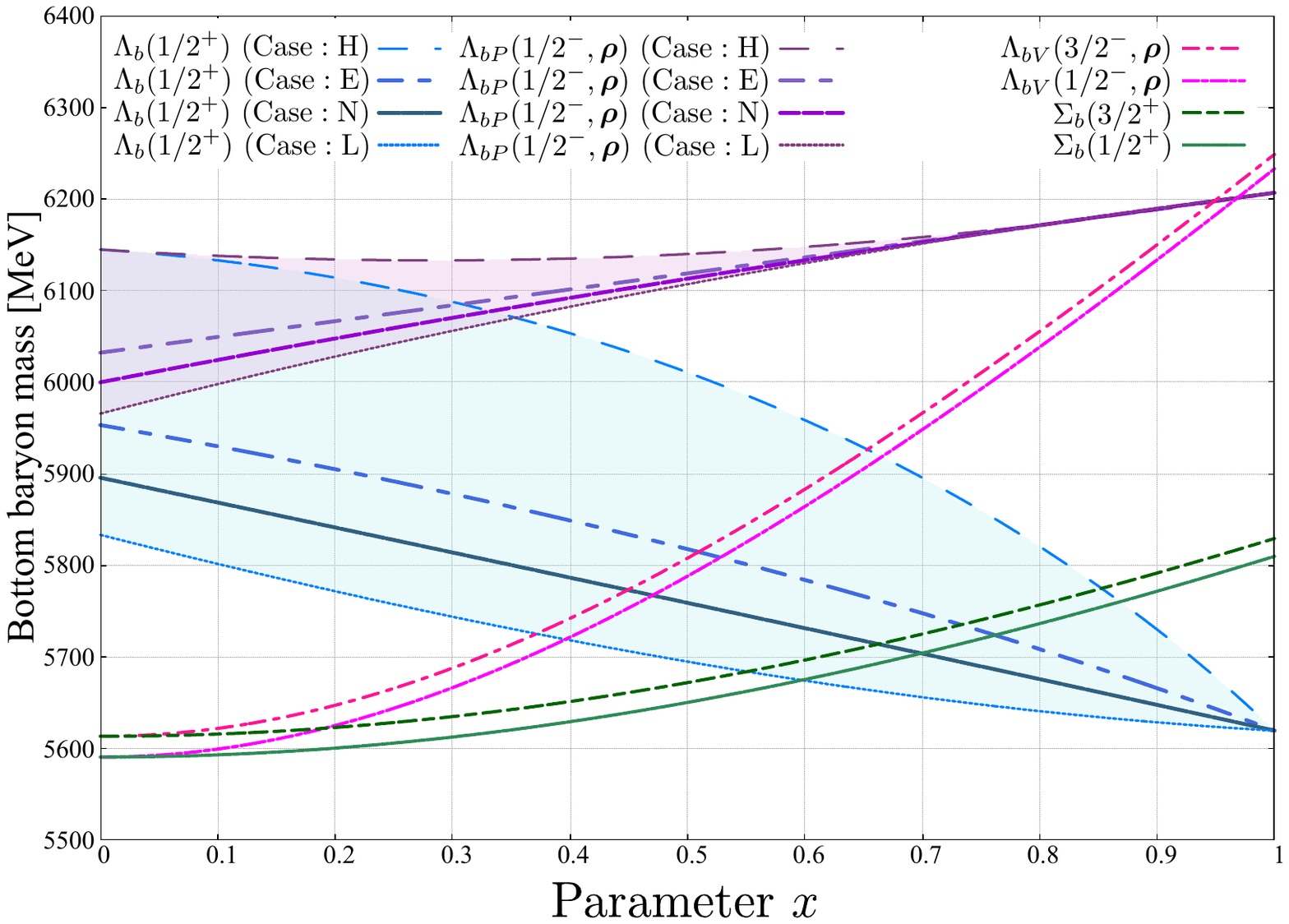}
  \includegraphics[clip,width=1.0\columnwidth]{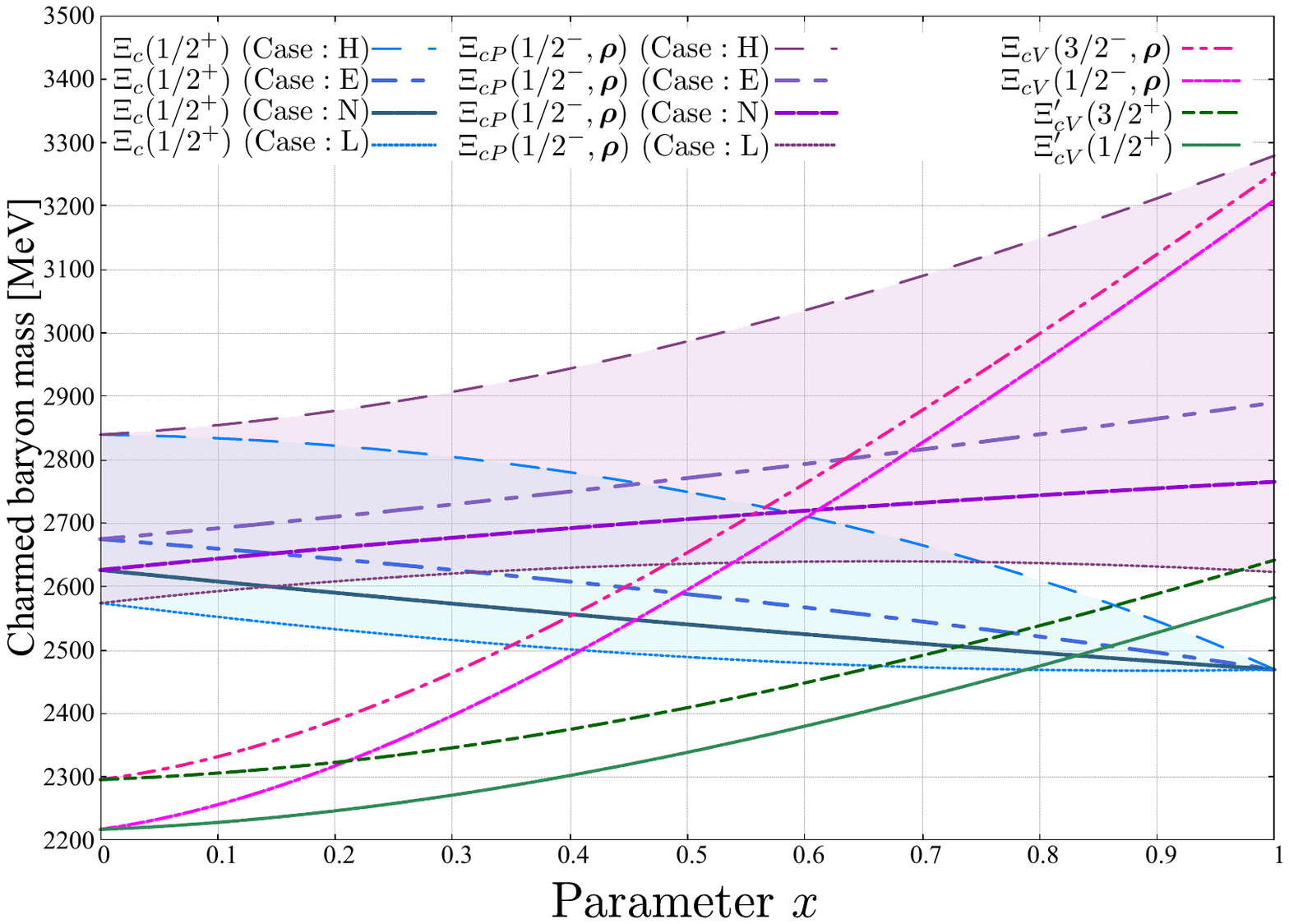}
  \includegraphics[clip,width=1.0\columnwidth]{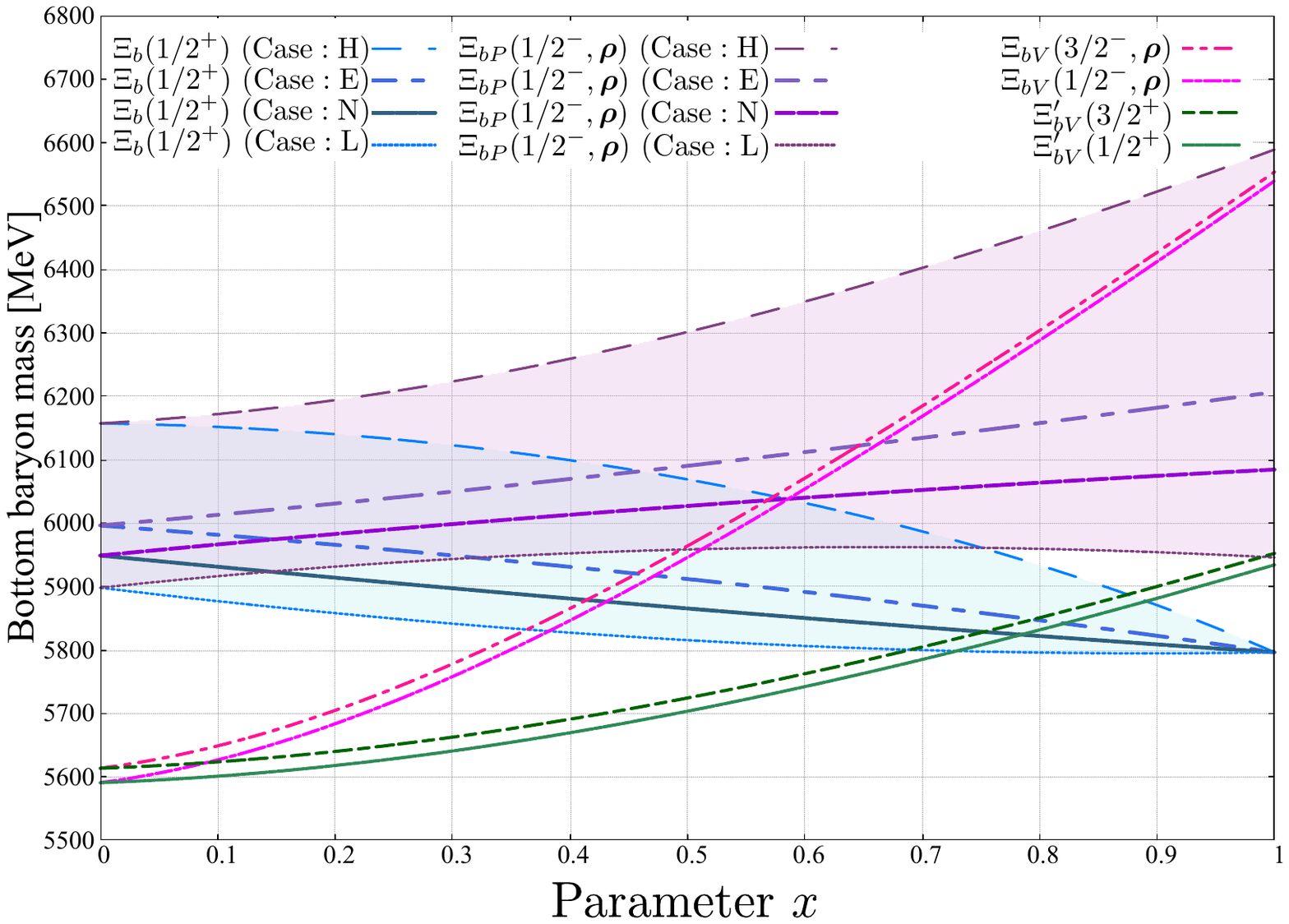}
  \caption{
Dependence of masses of singly heavy baryons, $cnn$ (left upper), $bnn$ (right upper), $csn$ (left lower), and $bsn$ (right lower) on the chiral symmetry breaking parameter $x$, obtained from the parameter choices, Case:L, Case:N, Case:E, and Case:H.
Cyan lines: the ground-state $\Lambda_Q(1/2^+)$ and $\Xi_Q(1/2^+)$.
Purple lines: the $\rho$-mode excited $\Lambda_{QP}(1/2^-,\bm{\rho})$ and $\Xi_{QP}(1/2^-,\bm{\rho})$.
Green lines: the ground-state $\Sigma_{Q}(1/2^+,3/2^+)$ and $\Xi_{Q}^\prime (1/2^+,3/2^+)$.
Magenta lines: the $\rho$-mode excited $\Lambda_{QV}(1/2^-,3/2^-,\bm{\rho})$ and $\Xi_{QV}(1/2^-,3/2^-,\bm{\rho})$.
The cyan and purple shaded regions designate the possible mass ranges.}
 \label{figure_Qnn}
\end{figure*}

As discussed in Sec.~\ref{Sec:restoration}, under chiral-symmetry restoration, the diquark masses are modified, which also changes the heavy-baryon masses.
We calculate the $x$-dependences of the heavy-baryon masses with the diquark--heavy-quark potential model.
The upper two panels in Fig.~\ref{figure_Qnn} show the results for the $x$-dependent masses of nonstrange singly heavy baryons, $\Lambda_{Q}(1/2^+)$, $\Lambda_{QP}(1/2^-,\bm{\rho})$, $\Sigma_{Q}(1/2^+,3/2^+)$, and $\Lambda_{QV}(1/2^-,3/2^-,\bm{\rho})$.
The masses of these baryons reflect the modification of the nonstrange diquark masses shown in Fig.~\ref{figure_Lagadd3}.
We find that the chiral partners of $\Lambda_{Q}(1/2^+)$ and $\Lambda_{QP}(1/2^-,\bm{\rho})$ approach each other by chiral restoration ($x \to 0$), but the mass difference remains at $x=0$ due to the effect of the axial anomaly, except for Case H.
On the other hand, the chiral partners of $\Sigma_{Q}(1/2^+,3/2^+)$ and $\Lambda_{QV}(1/2^-,3/2^-,\bm{\rho})$ completely degenerate at $x=0$.

The lower two panels in Fig.~\ref{figure_Qnn} show the results for the $x$-dependent masses of $\Xi_{Q}(1/2^+)$, $\Xi_{QP}(1/2^-,\bm{\rho})$, $\Xi^\prime_{Q}(1/2^+,3/2^+)$, and $\Xi_{QV}(1/2^-,3/2^-,\bm{\rho})$.
The masses of these baryons are affected by the singly strange diquark masses shown in Fig.~\ref{figure_Lagadd3s}.
Contrary to the nonstrange case, we can see that all the chiral partners are degenerate at $x=0$.

In addition, we comment on the lowest-state inversion.
Compared to the crossing point $0.5661< x_{AS} <0.8580$ between the nonstrange S- and A-diquark masses shown in Fig.~\ref{figure_Lagadd3}, the crossing point between $\Lambda_{Q}(1/2^+)$ and $\Sigma_{Q}(1/2^+)$ is shifted to the larger-$x$ side, while that between $\Lambda_{Q}(1/2^+)$ and $\Sigma_{Q}(3/2^+)$ is shifted to the smaller side.
Thus, the position of the crossing point is affected by the spin splitting of baryons.
However, this shift is suppressed in the bottom sector compared to the charm sector and should vanish in the heavy-quark limit.

\section{Decay width of $\Sigma_Q^{(*)} \rightarrow \Lambda_Q+\pi^\pm$} \label{Sec:decay}

As shown in Fig.~\ref{figure_Qnn}, we have predicted that the mass difference between $\Lambda_Q(1/2^+)$ and $\Sigma_Q(1/2^+,3/2^+)$ changes as $x\to 0$: $\Lambda_Q(1/2^+)$ becomes heavier than $\Sigma_Q(1/2^+,3/2^+)$ in the small $x$ region.
From this finding, we expect that such a change also modifies the property of decays. 
Here, we consider the strong decays, $\Sigma_Q^{(*)}\to \Lambda_Q + \pi^\pm$ 
where $\Sigma_Q$ denotes the spin 1/2 and isospin 1 baryon, and $\Sigma^*_Q$ is the one with spin 3/2 and isospin 1.

Using the diquark-heavy quark picture for singly heavy baryons in our previous work \cite{Kim:2022pyq},
the decay widths of $\Sigma_Q^{(*)}\rightarrow \Lambda_Q + \pi^\pm$ are written as
\begin{widetext}
\begin{eqnarray}
&&\Gamma[\Sigma_Q \rightarrow \Lambda_Q \pi]
=\frac{|\sqrt{2}G_1|^2 |\vec{p}_\pi| E_{\Lambda_Q}}{216\pi M_{\Sigma_Q}}
\left\{
 \frac{|\vec{p}_{\pi}|^2(3+\upsilon_A^2) + 3E_\pi^2\upsilon_A^2}{3E_A E_S}
 \right\}
 |\langle\phi_{S}|\phi_{A, J} \rangle|^2 , \nonumber \\
 &&|\vec{p}_\pi|=\frac{\sqrt{\left[M_{\Sigma_Q}^2-(M_{\Lambda_Q}+M_\pi)^2\right]\left[M_{\Sigma_Q}^2-(M_{\Lambda_Q}-M_\pi)^2\right]}}{2M_{\Sigma_Q}}~,\\
 &&M_A^2 \upsilon_A^2 \equiv \langle p^2 \rangle \equiv \int \frac{d^3p}{(2\pi)^3} |\vec{p}|^2 |\phi_{A,J}(\vec{p})|^2,
 \nonumber
\label{width_final}
\end{eqnarray}
\end{widetext}
where the subscripts $A$ and $S$ denote the types of constituent diquarks, and $\phi_S$ and $\phi_{A,J}$ are the wave functions of the baryons, $\Lambda_Q$ and $\Sigma_Q^{(*)}$, as a bound state of $S$ and $A$ diquarks with the heavy quark $Q$ 
in the diquark--heavy-quark model.
The subscript $J$ in $\phi_{A,J}$ stands for the total spin $1/2$ or $3/2$ of $\Sigma_Q^{(*)}$ baryon.
The wave functions, $\phi_S$ and $\phi_{A,J}$, are numerically obtained by solving the two-body Schr\"odinger equation for the diquark--heavy-quark system with the Y-potential in Ref.~\cite{Kim:2021ywp}.
As the parameters, we input the hadron masses, $M_{\Sigma_Q}$, $M_{\Lambda_Q}$, and $M_{\pi}$, the hadron energies (in the rest frame of $\Sigma_Q^{(*)}$), $E_{\Lambda_Q} = \sqrt{M_{\Lambda_Q}^2+|\vec{p}_\pi|^2}$ and $E_{\pi}=\sqrt{M_\pi^2+|\vec{p}_\pi|^2}$, and the diquark energies
$E_A=\sqrt{M_A^2+M_A^2v_A^2}$ and $E_S=\sqrt{M_S^2+M_A^2v_A^2 + |\vec{p}_\pi|^2}$ with the diquark masses, $M_A$ and $M_S$.

\begin{figure*}[t!]
  \includegraphics[clip,width=1.0\columnwidth]{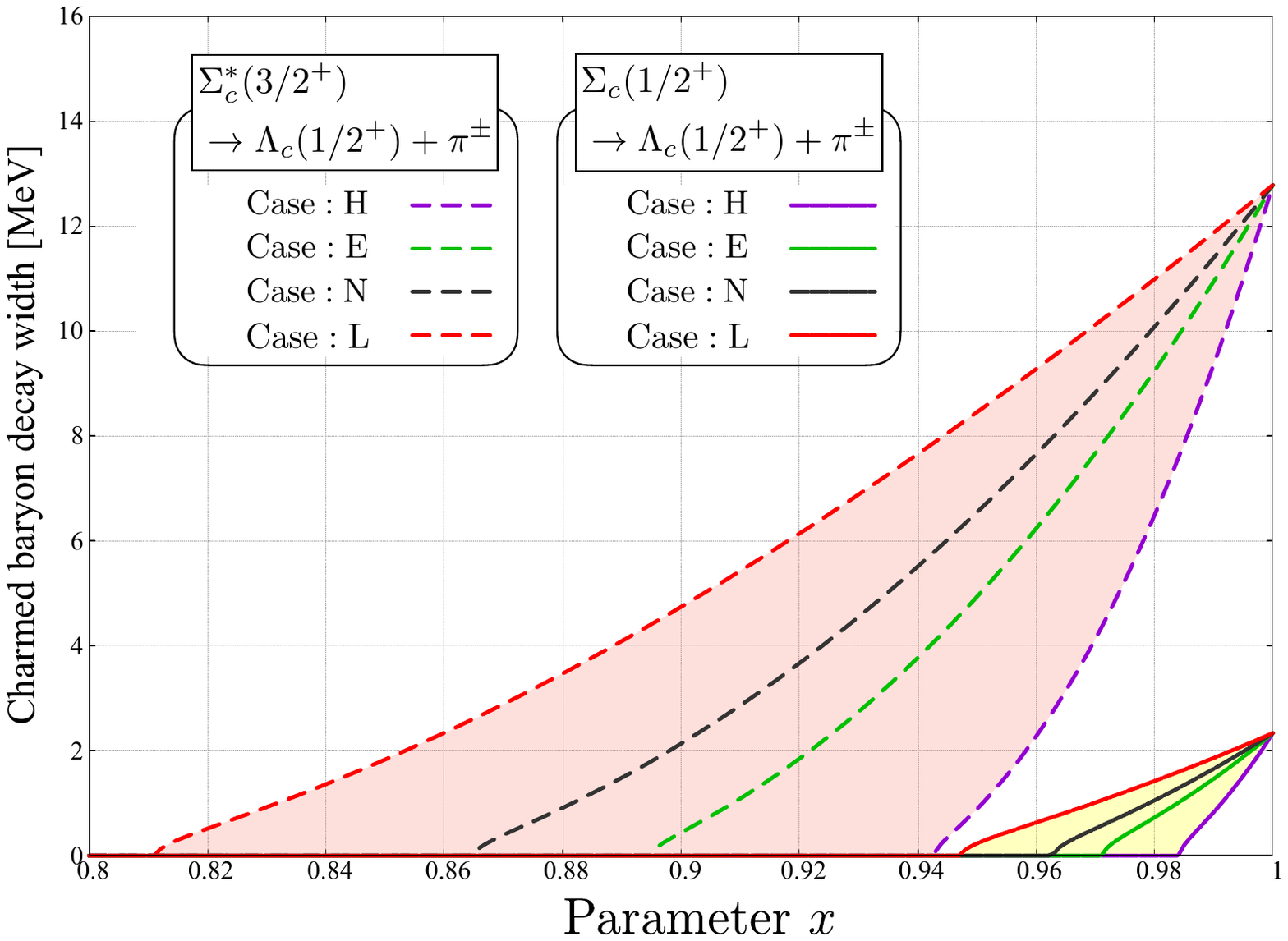}
  \includegraphics[clip,width=1.0\columnwidth]{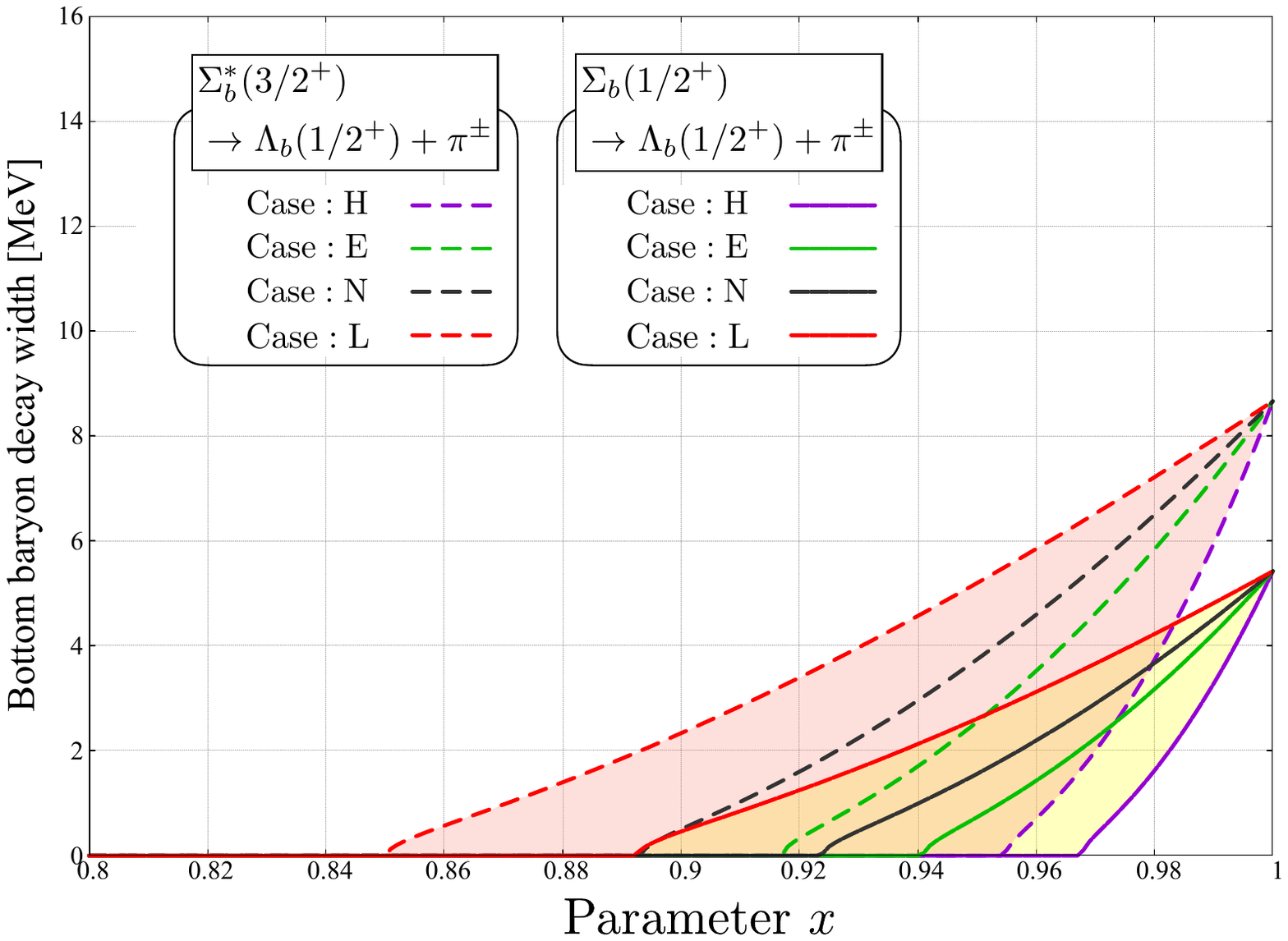}
  \caption{Dependence of the decay widths of $\Sigma_Q^{(*)} \rightarrow \Lambda_Q+\pi^\pm$ reactions on the chiral symmetry breaking parameter $x$ (see Figs.~3 and 4 in Ref.~\cite{Kim:2022pyq} as the previous study).
The solid and dotted lines stand for $\Sigma_Q(1/2^+) \rightarrow \Lambda_Q(1/2^+)+\pi^\pm$ and $\Sigma_Q^*(3/2^+) \rightarrow \Lambda_Q(1/2^+)+\pi^\pm$, respectively.
The shaded regions are filled as the range between Case:L (red) and Case:H (purple).
Case:N (black) and Case:E (green) are drawn as the colored lines.
}
 \label{figure_width}
\end{figure*}

The coupling constant $G_1$ is defined by
\begin{eqnarray}
&& G_1=g_1+\left(\frac{m_s}{g_s f_\pi}+x\frac{f_s}{f_\pi} \right)g_2,
\end{eqnarray} 
and $g_{1,2}$ are the bare couplings given in the Lagrangian for the S-V coupling \cite{Kim:2022pyq}
\begin{eqnarray}
&&\lag_{S-V}=g_1\epsilon_{ijk}
\left[
d^\mu_{ni} (\partial_\mu \Sigma^\dag)_{jn} d^\dag_{R,k}
+d^\mu_{in} (\partial_\mu \Sigma)_{jn} d^\dag_{L,k}
 \right] \nonumber\\
&&
+\frac{g_2}{f_\pi}\epsilon_{ijk}
\left[
d^\mu_{in} \{\Sigma_{jn}(\partial_\mu \Sigma)_{km}-(\partial_\mu \Sigma)_{jn}\Sigma_{km}\} d^\dag_{R,m}\right. \nonumber\\
&&\left.
+d^\mu_{ni} \{\Sigma^\dag_{jn}(\partial_\mu \Sigma^\dag)_{km}-(\partial_\mu \Sigma^\dag)_{jn}\Sigma^\dag_{km}\} d^\dag_{L,m} 
 \right].
 \label{SVLag}
\end{eqnarray}
Note that the bare coupling constants $g_{1,2}$ are independent of $x$, while $G_1$ depends on $x$ through 
the second term of the Lagrangian Eq.~(\ref{SVLag}).
In the numerical calculation, we use the values of $g_{1,2}$ determined in Ref.~\cite{Kim:2022pyq} 
separately for each heavy quark sector ($Q=c,b$) as
\begin{align}
&(g^c_1, g^c_2)=(30.56,  -3.66    ), \nonumber\\
&(g^b_1, g^b_2)=(33.28,  -6.12    ).
\label{g1g2value}
\end{align}
In Table~\ref{decay_exp}, we show the comparison of our prediction at $x=1$ and the corresponding experimental data.

\begin{table}[tb!]
  \centering
    \caption{Decay widths of $\Sigma_Q^{(*)}\rightarrow\Lambda_Q + \pi^{\pm}$ strong decay in the ordinary vacuum ($x=1$) compared with the experimental data. For the decay width, (cal.) and (exp.) represent our calculated results~\cite{Kim:2022pyq} and experimental data in PDG~\cite{ParticleDataGroup:2024cfk}, respectively.
    }
  \begin{tabular}{ c  c  c  c  c  } \hline\hline
       \multicolumn{1}{c}{}
       &\multicolumn{1}{c}{}
       &\multicolumn{1}{c}{}
       &\multicolumn{2}{c}{Decay width (MeV)} \\
\cline{4-5}
       \multicolumn{1}{l}{Baryon}
       &\multicolumn{1}{c}{$J^P$}
       &\multicolumn{1}{c}{Decay mode}
       &\multicolumn{1}{c}{(cal.)}
       &\multicolumn{1}{c}{(exp.)}    \\ \hline       
       
       \hline

$\Sigma_c^{++}(2455)$&$1/2^+$	&$\Sigma_c^{++} \rightarrow \Lambda_c \pi^+$		&2.336	&1.89  	 \\ 
$\Sigma_c^{0}(2455) $&$1/2^+$	&$\Sigma_c^{0} \rightarrow \Lambda_c \pi^-$		&2.336	&1.83  	\\ \hline
$\Sigma_c^{++}(2520) $&$3/2^+$	&$\Sigma_c^{*++} \rightarrow \Lambda_c \pi^+$	&12.78	&14.78	\\ 
$\Sigma_c^{0}(2520) $&$3/2^+$	&$\Sigma_c^{*0} \rightarrow \Lambda_c \pi^-$		&12.78	&15.3  	\\ \hline

\hline
$\Sigma_b^{+} $&$1/2^+$	&$\Sigma_b^+ \rightarrow \Lambda_b\pi^+$		&5.423	&5.0  	 \\ 
$\Sigma_b^{-} $&$1/2^+$	&$\Sigma_b^- \rightarrow \Lambda_b\pi^-$		&5.423	&5.3  	\\ \hline
$\Sigma_b^{*+} $&$3/2^+$	&$\Sigma_b^{*+} \rightarrow \Lambda_b\pi^+$	&8.660	&9.4  	\\ 
$\Sigma_b^{*-} $&$3/2^+$	&$\Sigma_b^{*-} \rightarrow \Lambda_b\pi^-$	&8.660	&10.4	\\ \hline\hline
  \end{tabular}
  \label{decay_exp}
\end{table}

\begin{table*}[t!]
  \centering
\caption{Numerical values of the crossing point $x_{AS}$ where the nonstrange A- and S-diquark masses are degenerate (from Fig.~\ref{figure_Lagadd3}) and the threshold $x_{\rm th}$ where $\Sigma_Q^{(*)}\rightarrow \Lambda_Q + \pi^\pm$ decay is hindered (from Fig.~\ref{figure_width}).}
  \begin{tabular}{l c  c cccc} \hline\hline 

&
&
&\multicolumn{4}{c}{Threshold $x_\mathrm{th}$ for $\Sigma_Q^{(*)}$ decay}   
\\
\cline{4-7}
Cases
&$M_0$ (MeV)
& Crossing point $x_{AS}$
&$\Sigma_c(1/2^+)$
&$\Sigma_c^*(3/2^+)$ 
&$\Sigma_b(1/2^+)$
&$\Sigma_b^*(3/2^+) $
\\ \hline

\hline
Case:L		&2623		
&0.5661  &0.947&0.811	&0.892&0.850   \\ 
Case:N	&2765		
&0.6747   &0.962&0.864	&0.923&0.893  \\
Case:E		&2890		
&0.7425   &0.971&0.895 &0.940&0.917\\
Case:H		&3279		
&0.8580 &0.984&0.942	&0.967&0.954 \\ 
   \hline\hline
  \end{tabular}

  \label{AScross}
\end{table*}

In Fig.~\ref{figure_width}, we show the results for the $x$ dependence of decay widths for the charm and bottom sectors.%
\footnote{The result of Case:N is almost the same as Figs.~3 and 4 in Ref.~\cite{Kim:2022pyq}, but slightly different due to the error in the $x$-dependence of diquark mass formulas.
The correct result is in Fig.~\ref{figure_width}.}
Here, we have assumed that the pion mass is independent of $x$ 
because we expect that the pion mass may not change much in the region, $0.8 \le x \le 1$.

From Fig.~\ref{figure_width}, we can see that the decay widths of $\Sigma_Q^{(*)}\rightarrow \Lambda_Q + \pi^\pm$ are suppressed as $x$ decreases, and the decay is eventually forbidden.
Such a {\it missing/suppressed decay}~\cite{Kim:2022pyq} can be signals of partially restored chiral symmetry.
Here, we define the threshold $x_{\rm th}$ as the value of $x$ where the decay is hindered.
Our numerical results of $x_{\rm th}$ are summarized in Table \ref{AScross}. 
We find that the threshold $x_{\rm th}$ is also moving toward $x=1$ as the parameter $M_0$ increases.
This is similar to the behavior of the crossing points $x_{AS}$ between the A and S diquark masses, but in all the cases, this threshold is larger than the crossing point, i.e., $x_{AS}<x_{\rm th}$.
Our results indicate that about 20\% of partial restoration of chiral symmetry forbids these strong decays. 
Therefore, by experimentally measuring decay widths of heavy baryons in (partially) chiral-symmetry restored medium, we can determine the parameters of our effective Lagrangian.

\section{Summary} \label{Sec:summary}

In summary, we have constructed a new chiral effective Lagrangian~(\ref{sLag_plus}) for the S and P diquarks, where the chiral $SU(3)_R\times SU(3)_L$ and $U(1)_A$ symmetric term with the coefficient $\mu_0^2$ is newly added, which was omitted in our previous study~\cite{CETdiquark,scalar}.

From this Lagrangian, the new diquark-mass formulas are obtained as Eqs.~(\ref{Smass1})--(\ref{Smass4}).
We have also investigated the behaviors of the diquark masses when the spontaneous chiral symmetry breaking is partially or totally restored.
The new mass formulas for the S/P diquarks under a chiral-symmetry restoration are given as Eqs.~(\ref{newSmassx}), (\ref{newPmassx}), (\ref{Ssnmassx}), and (\ref{Psnmassx}) (see Figs.~\ref{figure_Lagadd3} and \ref{figure_Lagadd3s}).
Based on the diquark--heavy-quark two-body picture for singly heavy baryons, by inputting the diquark masses, we have predicted the spectra in vacuum (Figs.~\ref{figure_Cqqspectrum} and \ref{figure_Bqqspectrum}) and the behaviors of the baryon masses (Fig.~\ref{figure_Qnn}) and decay widths (Fig.~\ref{figure_width}) under a chiral-symmetry restoration.

Among the parameter sets extended by taking into account the $\mu_0^2$ term (see Table~\ref{Xirho_parameter}), in the future we have to determine which parameter set is reasonable in nature or in QCD.
The true parameter set will be determined by examining the following characteristic phenomena:
\begin{enumerate}
\item[(i)] {\it Inverse/normal mass hierarchy}---In vacuum (i.e., at zero temperature/density), the mass ordering of nonstrange/strange P diquarks has to be established.
We find that, depending on the parameter set, either the inverse mass ordering~\cite{CETdiquark,scalar}, $M_{ud}(0^-) >M_{ns}(0^-)$, or the normal mass ordering, $M_{ud}(0^-) <M_{ns}(0^-)$, is allowed.
This ordering is also reflected in the mass ordering of singly heavy baryons including the P diquark, $\Lambda_{QP}(1/2^-,{\bm \rho})$ and $\Xi_{QP}(1/2^-,{\bm \rho})$.
Therefore, to measure this mass ordering by lattice QCD simulations or experiments is crucial to determine the parameters of our Lagrangian.
\item[(ii)] {\it Lowest-state inversion}---In vacuum, the S diquark is the lightest of all diquarks.
In our model, when the spontaneous chiral symmetry breaking is sufficiently weakened at finite temperature/density, the mass ordering of the S and A diquark is inverted~\cite{Kim:2021ywp}: $M(0^+) >M(1^+)$.
This inversion occurs in all the parameter sets, but the position of crossing point $x_{AS}$ depends on the parameters.
Furthermore, this inversion is also reflected in the mass ordering of singly heavy baryons including the S or A diquark, e.g., $\Lambda_Q(1/2^+) >\Sigma_Q(1/2^+)$.
Therefore, which state is the lowest should be examined by lattice QCD simulations or experiments. 
\item[(iii)] {\it Missing/Suppressed decays}---Due to the lowest-state inversion of singly heavy baryons, the decay width of $\Sigma_Q^{(*)} (1/2^+,3/2^+)\rightarrow \Lambda_Q (1/2^+)+ \pi^\pm$ is suppressed and eventually hindered~\cite{Kim:2022pyq}.
Its threshold $x_\mathrm{th}$ depends a little on the parameter set.
Such behaviors of decay widths will be a good signal in each context of the partially restored chiral symmetry, the diquark correlations inside heavy baryons, or the determination of Lagrangian parameters.
\item[(iv)] {\it Nondegeneracy of chiral partners}---Even when the spontaneous chiral symmetry breaking is fully restored, a mass splitting between chiral partners of the nonstrange S and P diquarks, or $\Lambda_{Q}(1/2^+)$ and $\Lambda_{QP}(1/2^-,{\bm \rho})$, may remain.
Within our model, this splitting is induced by the $U(1)_A$ anomaly characterized by $m_{S1}^2 \neq 0$~\cite{Kim:2021ywp}, and if $m_{S1}^2 \sim 0$, it disappears.
Therefore, the measurement of this splitting is important for determining not only the presence or absence of the $m_{S1}^2$ term but also its magnitude.
\end{enumerate}

Finally, we comment on tetraquarks.
Similar to the diquark--heavy-quark picture for singly heavy baryons ($Qqq$), doubly heavy tetraquarks ($QQqq$) can be described as one diquark and two heavy quarks.
Based on this picture, the above phenomena appear also in doubly heavy tetraquarks~\cite{Kim:2022mpa} and will be useful for further understanding diquarks inside hadrons.


\section*{Acknowledgments}
This work was supported by Grants-in-Aid for Scientific Research,
Grants No. JP20K14476, No. JP21H00132, No. JP23K03427, and No. JP24K07034
from Japan Society for the Promotion of Science (JSPS).

\section*{Data availability}
No data were created or analyzed in this study.

\bibliography{reference-Supdate}
\end{document}